\documentclass[12pt,a4paper]{article}
\usepackage[dvips]{graphicx}
\usepackage{cite}
\newcommand{\rs}     {\ensuremath{\sqrt{s}}}
\newcommand{\rsp}    {\ensuremath{\sqrt{s'}}}

\newcommand{\as}     {\ensuremath{\alpha_s}}
\newcommand{\asmz}   {\ensuremath{\as(M_\mathrm{Z})}}
\newcommand{\assp}   {\ensuremath{\as(\sqrt{s'}})}

\newcommand{\Mz}     {\ensuremath{M_{\mathrm{Z}}}}
\newcommand{\Z}      {\ensuremath{\mathrm{Z}^0}}
\newcommand{\epem}   {\ensuremath{\mathrm{e}^+\mathrm{e}^-}}
\newcommand{\qqbar}  {\ensuremath{\mathrm{q\bar{q}}}}
\newcommand{\elm}     {\ensuremath{\mathrm{e}^-}}
\newcommand{\Jetset} {\texttt{JETSET}}
\newcommand{\Herwig} {\texttt{HERWIG}}
\newcommand{\Ariadne}{\texttt{ARIADNE}}
\newcommand{\Phojet} {\texttt{PHOJET}}
\newcommand{\Vermaseren}{\texttt{VERMASEREN}}
\newcommand{\KoralZ} {\texttt{KORALZ}}
\newcommand{\mnch}   {\ensuremath{\langle n_{\mathrm{ch}}\rangle}}
\newcommand{\bm}[1]  {\mbox{\boldmath\ensuremath{#1}}}
\newcommand{\mh}     {\ensuremath{M_H}}
\newcommand{\thr}    {\ensuremath{1-T}}
\newcommand{\bt}     {\ensuremath{B_T}}
\newcommand{\bw}     {\ensuremath{B_W}}
\newcommand{\xmu} {\ensuremath{x_{\mu}}}

\newcommand{\opalh}{\begin{center} \huge $\mathbf{OPAL}$ \end{center}}

\newcommand{\Piso}{\ensuremath{P_{\mathrm{iso}}}}
\newcommand{\Eiso}{\ensuremath{E_{\mathrm{iso}}}}
\newcommand{\Aiso}{\ensuremath{\alpha_{\mathrm{iso}}}}
\newcommand{\chisq}{\ensuremath{\chi^2}}
\newcommand{\chisqd}{\ensuremath{\chi^2/\mathrm{d.o.f.}}}

\newcommand{\lrqqg} {\ensuremath{{\cal L}_{\qqbar\gamma}}}
\newcommand{\lqqg}  {\ensuremath{L_{\qqbar\gamma}}}
\newcommand{\erf}   {\ensuremath{\mathrm{erf}}}

\begin{document}

\begin{titlepage}
\begin{center}
{\large EUROPEAN ORGANIZATION FOR NUCLEAR RESEARCH}
\end{center}
\bigskip
\begin{flushright}
       CERN-PH-EP/2007-032   \\ 5 September 2007
\end{flushright}
\bigskip\bigskip\bigskip\bigskip\bigskip
\begin{center}
\begin{boldmath}
 {\huge\bf Measurement of \as\ with Radiative Hadronic Events }
\end{boldmath}
\end{center}
\bigskip\bigskip
\begin{center}
{\LARGE The OPAL Collaboration}
\end{center}
\bigskip\bigskip\bigskip
\begin{center}
{\large Abstract}
\end{center}

Hadronic final states with a hard isolated photon are studied using
data taken at centre-of-mass energies around the mass of the \Z\ boson
with the OPAL detector at LEP.  The strong coupling \as\ is extracted
by comparing data and QCD predictions for event shape observables at
average reduced centre-of-mass energies ranging from 24~GeV to 78~GeV,
and the energy dependence of \as\ is studied.  Our results are
consistent with the running of \as\ as predicted by QCD and show that
within the uncertainties of our analysis event shapes in hadronic \Z\
decays with hard and isolated photon radiation can be described by QCD
at reduced centre-of-mass energies.  Combining all values from
different event shape observables and energies gives $\asmz=0.1182\pm
0.0015(\mathrm{stat.})\pm 0.0101(\mathrm{syst.})$.

\bigskip\bigskip\bigskip\bigskip\bigskip\bigskip

\begin{center}
{\large (submitted to Eur. Phys. J. C)}
\end{center}

%\begin{center}
% {\Large\bf Dispatch Draft Version 1.2, \today } \\
% Author: D.~Toya \\
% editing co-author: S.~Kluth \\
% Editorial Board: M.~Ford, T.~Kawamoto, S.~Kluth, S.~Lloyd \\
% Please submit comments by 16 June 2007.\\
%The author and the editorial board agreed that a member of the
%EB, S.~Kluth, makes the neccessary changes to the draft
%in response to comments by the collaboration in consultation
%with the author and the EB.
%\end{center}

\end{titlepage}

\begin{center}{\Large        The OPAL Collaboration
}\end{center}\bigskip
\begin{center}{
%begin authorlist PLEASE DO NOT DELETE THIS COMMENT
G.\thinspace Abbiendi$^{  2}$,
C.\thinspace Ainsley$^{  5}$,
P.F.\thinspace {\AA}kesson$^{  7}$,
G.\thinspace Alexander$^{ 21}$,
G.\thinspace Anagnostou$^{  1}$,
K.J.\thinspace Anderson$^{  8}$,
S.\thinspace Asai$^{ 22}$,
D.\thinspace Axen$^{ 26}$,
I.\thinspace Bailey$^{ 25}$,
E.\thinspace Barberio$^{  7,   p}$
% ...
%end authorlist PLEASE DO NOT DELETE THIS COMMENT
}\end{center}\bigskip
\bigskip
%begin institutes
$^{  1}$School of Physics and Astronomy, University of Birmingham,
Birmingham B15 2TT, UK
\newline
$^{  2}$Dipartimento di Fisica dell' Universit\`a di Bologna and INFN,
I-40126 Bologna, Italy
\newline
% ...
$^{  5}$Cavendish Laboratory, Cambridge CB3 0HE, UK
\newline
% ...
$^{  7}$CERN, European Organisation for Nuclear Research,
CH-1211 Geneva 23, Switzerland
\newline
$^{  8}$Enrico Fermi Institute and Department of Physics,
University of Chicago, Chicago IL 60637, USA
\newline
% ...
$^{ 21}$Department of Physics and Astronomy, Tel Aviv University,
Tel Aviv 69978, Israel
\newline
$^{ 22}$International Centre for Elementary Particle Physics and
Department of Physics, University of Tokyo, Tokyo 113-0033, and
Kobe University, Kobe 657-8501, Japan
\newline
% ...
$^{ 25}$University of Victoria, Department of Physics, P O Box 3055,
Victoria BC V8W 3P6, Canada
\newline
$^{ 26}$University of British Columbia, Department of Physics,
Vancouver BC V6T 1Z1, Canada
\newline
% ...
$^{  p}$ now at The University of Melbourne, Victoria, Australia
\newline
% ...
%end notes
\bigskip

\clearpage

%%%%%%%%%%%%%%%%%%%%%%%%%%%
\section{Introduction}
\label{Introduction}
%%%%%%%%%%%%%%%%%%%%%%%%%%%

In the theory of strong interactions, Quantum Chromodynamics
(QCD)~\cite{Fritzsch73,Gross:1973id,Politzer:1973fx}, the strong
coupling constant \as\ is predicted to decrease for high energy or
short distance reactions: a phenomenon known as asymptotic freedom.
Values of \as\ at different energy scales have been measured at
PETRA and LEP in \epem\ reactions with different centre-of-mass (cms)
energies ranging from 35 to 209 GeV and confirm the
prediction~\cite{OPALJade,l3290,aleph265,delphi327,OPALPR404,OPALPR414,jader4,kluth06}.

Assuming that photons emitted before or immediately after the $\Z$
production do not interfere with hard QCD processes, a measurement of
\as\ at the reduced cms energies, \rsp, of the hadronic system is
possible by using radiative multi-hadronic events, i.e.\
$\epem\rightarrow\qqbar\gamma$ events.

Most photons emitted from the incoming particles before the \Z\
production (initial state radiation, ISR) escape along the beam pipe
of the experiment.  Measurements of cross-sections for hadron
production with ISR have been presented by the KLOE and BaBar
collaborations~\cite{KLOE1,KLOE2,denig06,BABAR}.  In \epem\
annihilation to hadrons on the \Z\ peak isolated high energy photons
observed in the detector are mostly emitted by quarks produced in
hadronic \Z\ decays (final state radiation, FSR), because on the \Z\
peak ISR effects are suppressed.  Measurements of \as\ in hadronic
events with observed photons have been performed by the L3 and DELPHI
Collaborations~\cite{L3AsRad,DELPHIas}.  The DELPHI collaboration has
also measured the mean charged particle multiplicity $\mnch(s')$ using
FSR in~\cite{DELPHINchRad}.

When an energetic and isolated photon is emitted in the parton shower
the invariant mass of the recoiling parton system is taken to set the
energy scale for hard QCD processes such as gluon radiation. In parton
shower models~\cite{Jetset,Herwig,Ariadne} the invariant mass of an
intermediate parton or the transverse momentum of a parton branching
are used as ordering parameters for the parton shower development.  In
this picture an energetic and isolated photon must be produced at an
early stage of the shower evolution and therefore can be used to
deduce the scale for subsequent QCD processes.  The validity of this
method will be studied below using parton shower Monte Carlo programs.

Here we report on a measurement of \as\ from event shape observables
determined from the hadronic system in events with observed energetic
and isolated photons in the OPAL experiment.

\section{Analysis method}
\label{sec_method}

The reduced cms energy, \rsp, is defined by
$2E_{\mathrm{beam}}\sqrt{1-E_{\gamma}/E_{\mathrm{beam}}}$, where
$E_{\gamma}$ is the photon energy and $E_{\mathrm{beam}}$ is the beam
energy.  The flavour mixture of hadronic events in this analysis is
changed compared to non-radiative \Z\ decay events. The fraction of
up-type quarks is larger due to their larger electric charge.
However, since the strong interaction is blind to quark flavour in the
Standard Model, as e.g.\ demonstrated in~\cite{AsFlavDep}, the
difference is not taken into account.  The effects of massive b quarks
on hadronisation corrections are considered below as a systematic
uncertainty.

The determination of \as\ is based on measurements of event shape 
observables, which are calculated from all particles with momenta $p_i$
in an event:
\begin{description}
\item[Thrust \bm{T}.] The thrust $T$ is defined by the expression
  \begin{equation}
  T= \max_{\vec{\hat{n}}}\left(\frac{\sum_i|\vec{p}_i\cdot\vec{\hat{n}}|}
                    {\sum_i|\vec{p}_i|}\right)\;\;\;.
  \label{equ_thrust}
  \end{equation}
  The thrust axis $\vec{\hat{n}}_T$ is the direction $\vec{\hat{n}}$ which
  maximises the expression in parentheses.  A plane through the origin
  and perpendicular to $\vec{\hat{n}}_T$ divides the event into two
  hemispheres $H_1$ and $H_2$.
\item[Heavy Jet Mass \bm{\mh}.] The hemisphere 
invariant masses are calculated using  the particles
  in the two hemispheres $H_1$ and $H_2$.   We define
  \mh\
as the heavier mass, divided by $\rs$ .
\item[Jet Broadening variables \bm{\bt} and \bm{\bw}.] 
  These are defined by computing the quantity
  \begin{equation}
    B_k= \left(\frac{\sum_{i\in H_k}|\vec{p}_i\times\vec{\hat{n}}_T|}
                    {2\sum_i|\vec{p}_i|}\right)
  \label{equ_btbw}
  \end{equation}
  for each of the two event hemispheres, $H_k$,  defined above.
  The two observables are defined by
  \begin{equation}
    \bt= B_1+B_2\;\;\;\mathrm{and}\;\;\;\bw= \max(B_1,B_2)\;\;\;
  \end{equation}
  where \bt\ is the total and \bw\ is the wide jet broadening.
\item[C-parameter \bm{C}.] The linear momentum tensor 
$\Theta^{\alpha\beta}$ is defined by
\begin{equation}
   \Theta^{\alpha\beta} = 
  \frac{\sum_i\vec{p}_i^\alpha\vec{p}_i^\beta/|\vec{p}_i|}{\sum_j|\vec{p}_j|}
  \ , \ \alpha,\beta = 1,2,3.
\end{equation}
The three eigenvalues $\lambda_j$ of this tensor define $C$ with
\begin{equation}
   C = 3(\lambda_1\lambda_2+\lambda_2\lambda_3+\lambda_3\lambda_1).
\end{equation}
\item[Transition value $y_{23}^D$.] This observable is given by the 
value of $y_{cut}$ in the Durham algorithm where the number of jets in 
an event changes from two to three.
\end{description}

In order to verify that using hadronic \Z\ decays with hard and
isolated final state radiation allows one to extract \as\ at a
reduced scale \rsp\ we employ simulated events.  We use the Monte
Carlo simulation programs \Jetset\ version 7.4~\cite{Jetset},
\Herwig\ version 5.9~\cite{Herwig} and \Ariadne\ version
4.08~\cite{Ariadne}, which have different implementations of the parton
shower algorithms including simulation of FSR. One
sample contains hadronic \Z\ decays with FSR and ISR (375 k events)
while the other samples are generated at lower cms energies without
ISR (500 k events each).  

We consider the generated events after the parton shower has stopped
(parton-level) and calculate event shape observables using the
remaining partons.  The effective cms energy \rsp\ is calculated from
the parton four-momenta excluding any final state photons and the
events are boosted into the cms system of the partons.  The samples
are binned according to the energy $E_{\mathrm{FSR}}$ of any FSR in
intervals of 5~GeV width for $E_{\mathrm{FSR}} > 10$~GeV.

\begin{figure}[htb]
\opalh
\vspace{-0.30cm}
\centering
\includegraphics[width=\textwidth]{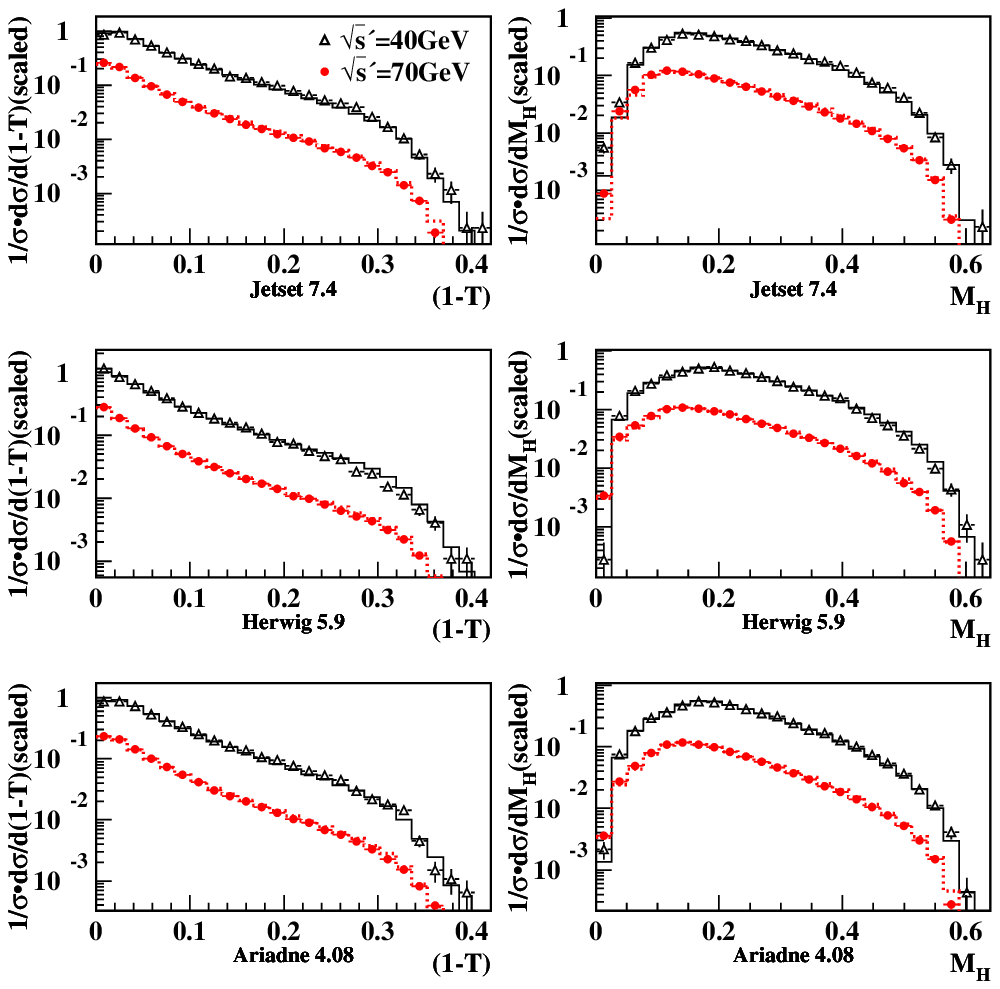}
\caption{The distributions of event shape observables \thr\ and 
  \mh\ for non-radiative events and radiative hadronic events from the
  Monte Carlo generators \Jetset, \Herwig\ and \Ariadne\ as indicated
  below the figures.  The triangles and points show distributions
  obtained from the \Z\ samples with FSR while the histograms show
  distributions from samples generated at lower energies as shown on
  the figure.  The open triangles and solid histogram (solid points
  and dashed histogram) in each figure correspond to $\rsp=40$
  (70)~GeV.}
\label{FigEvShpRadVSNRadDist}
\end{figure}

We observe good agreement between the corresponding distributions
obtained from the \Z\ sample with FSR and the lower energy samples.
For example, Figure~\ref{FigEvShpRadVSNRadDist} shows distributions of
the event shape observables \thr\ and \mh\ for two samples with
$\rsp=40$ and 70~GeV.  We conclude that within the approximations
made in the parton shower algorithms, hadronic \Z\ decays with hard and
isolated final state radiation can be used to extract measurements of
\as\ at reduced scales \rsp.

%%%%%%%%%%%%%%%%%%%%%%%%%%%
\section{The OPAL Detector and Event Simulation}
\label{DetectorEventSimulation}
%%%%%%%%%%%%%%%%%%%%%%%%%%%

The OPAL detector operated at the LEP \epem\ collider at CERN from
1989 to 2000.  A detailed description of the detector can be found
in~\cite{OpalDetector}.  We describe briefly the important parts of
the detector for this study.  In the OPAL coordinate system, the $x$
axis was horizontal and pointed approximately towards the centre of LEP,
the $y$ axis was normal to the $z$-$x$ plane , and the $z$ axis was in
the $\elm$ beam direction.  The polar angle, $\theta$, was measured
from the $z$ axis, and the azimuthal angle, $\phi$, from the $x$ axis
about the $z$ axis.

The central detector measured the momentum of charged particles and
consisted of a system of cylindrical drift chambers which lay within
an axial magnetic field of 0.435~T.  The momenta $p_{xy}$ of tracks in
the $x$-$y$ plane were measured with a precision of
$\sigma_p/p_{xy}=0.02\%\oplus 0.0015\cdot
p_{xy}[\mathrm{GeV}/c]$~\cite{OPALMomReso}.

The electromagnetic calorimeters completely covered the azimuthal
range for polar angles satisfying $|\cos{\theta}|<0.98$.  The barrel
electromagnetic calorimeter covered the polar angle range
$|\cos{\theta}|<0.82$, and consisted of a barrel of 9440 lead glass
blocks oriented so that they nearly pointed to the interaction region.
The two endcaps were each made of 1132 lead glass blocks, aligned
along the $z$-axis.  Each lead glass block in the barrel
electromagnetic calorimeter was 10$\times$10~cm$^2$ in cross section,
which corresponds to an angular region of approximately $40\times
40$~mrad$^2$.  The intrinsic energy resolution was $\sigma_E/E =
0.2\%\oplus 6.3\%/\sqrt{E[\mathrm{GeV}]}$~\cite{OpalDetector}.

Most electromagnetic showers were initiated before the lead glass
mainly because of the coil and pressure vessel in front of the
calorimeter.  An electromagnetic presampler made of limited streamer
tubes measured the shower position. The barrel presampler covered the
polar angle range $|\cos{\theta}|<0.81$ and its angular resolution for
photons was approximately 2~mrad.

\Jetset\ version 7.4 was used to simulate
$\epem\rightarrow\qqbar$ events, with
\Herwig\ version 5.9 and $\Ariadne$ version 4.08 used as
alternatives.  Parameters controlling the hadronisation of quarks and
gluons were tuned to OPAL LEP~1 data as described
in~\cite{TuneJetset,TuneHerwig}.  We used \Herwig\ version
5.9~\cite{Herwig}, \Phojet\ version 1.05c~\cite{Phojet1,Phojet2} and
\Vermaseren\ version 1.01~\cite{Vermaseren} for two-photon
interactions and \KoralZ\ version 4.02~\cite{KoralZ} for
$\epem\rightarrow\tau^+\tau^-$ events.  Generated events were processed
through a full simulation of the OPAL detector~\cite{GOPAL} and the
same event analysis chain was applied to the simulated events as to
the data. 4,000,000 fully simulated events were generated by \Jetset,
200,000 events, 1,000,000 events and 55,000 events were generated by
\Herwig, \Phojet\ and \Vermaseren\ while 800,000 events were generated
by \KoralZ.

%%%%%%%%%%%%%%%%%%%%%%%%%%
\section{Event Selection}
\label{EventSelection}
%%%%%%%%%%%%%%%%%%%%%%%%%%

\subsection{Hadronic Event Selection}
\label{MHSelection}

This study is based on a sample of 3 million hadronic \Z\ decays
selected as described in~\cite{LineShape} from the data accumulated
between 1992 and 1995 at cms energy of 91.2~GeV.  We required that the
central detector and the electromagnetic calorimeter were fully
operational.

For this study, we apply stringent cuts on tracks and clusters and
further cuts on hadronic events.  The clusters in the electromagnetic
calorimeter are required to have a minimum energy of 100~MeV in the
barrel and 250~MeV in the endcap.  Tracks are required to have
transverse momentum $p_T\ge 150~\mathrm{MeV}/c$ with respect to the
beam axis, at least 40 reconstructed points in the jet chamber, at the
point of closest approach a distance between the track and the nominal
vertex $d_0<2$~cm in the $r$-$\phi$ plane and $z_0<25$~cm in the $z$
direction.  We require at least five such tracks to reduce background
from $\epem\rightarrow\tau^+\tau^-$ and $\gamma\gamma\rightarrow\qqbar$
events.  The polar angle of the thrust axis is required to satisfy
$|\cos{\theta_T}|<0.9$, to ensure that events are well contained in
the OPAL detector.  After these cuts, a data sample of $2.4\times
10^6$ events remains.

\subsection{Isolated Photon Selection}

\subsubsection{Isolation Cuts}
\label{IsoEMSelection}

\begin{figure}[htb]
\opalh\vspace{-0.5cm}
\centering
\begin{tabular}{cc}
% Figures -2 fixed for lower case \theta or variable name C->S
\includegraphics[width=0.4\textwidth]{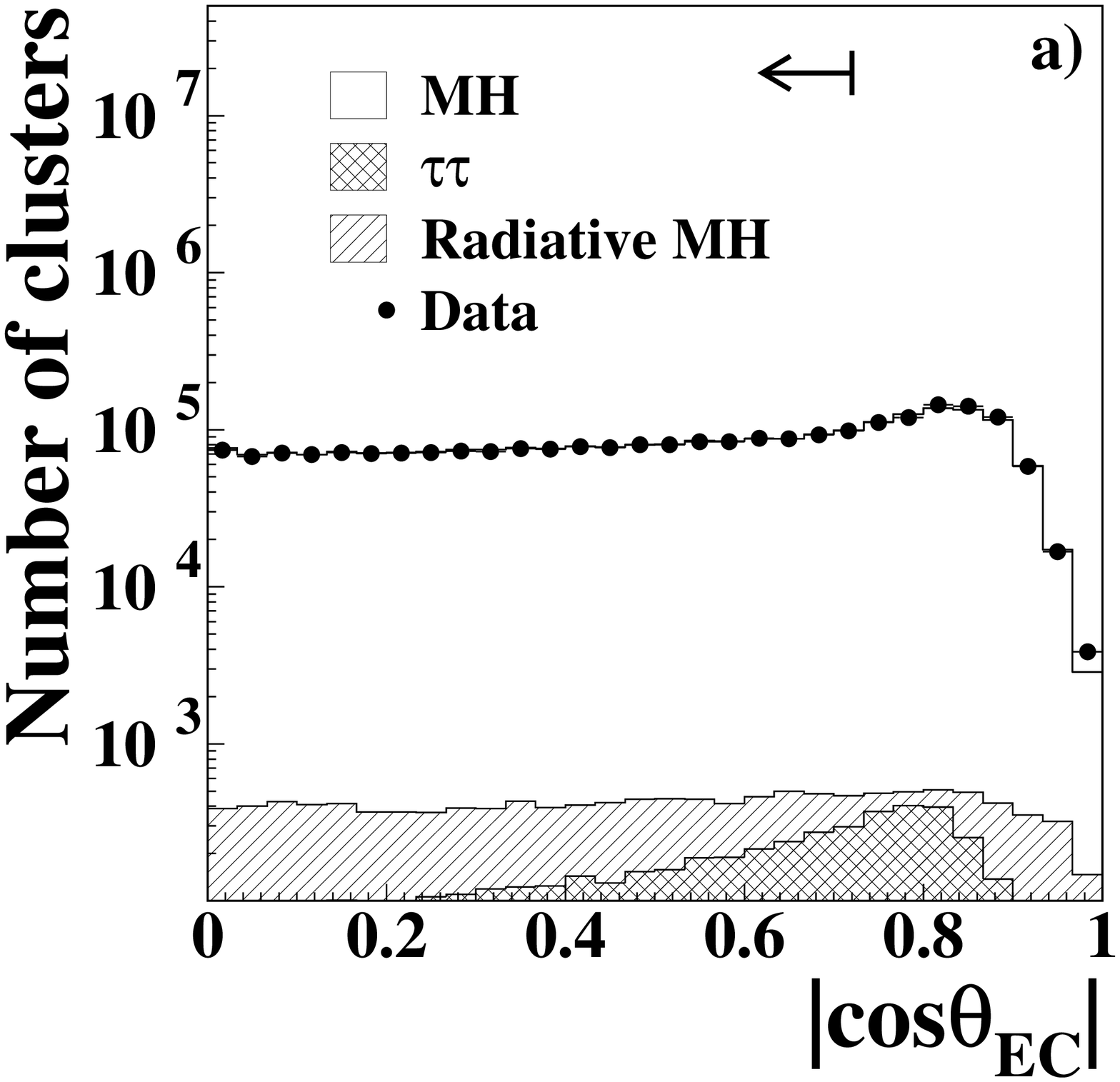} &
\includegraphics[width=0.4\textwidth]{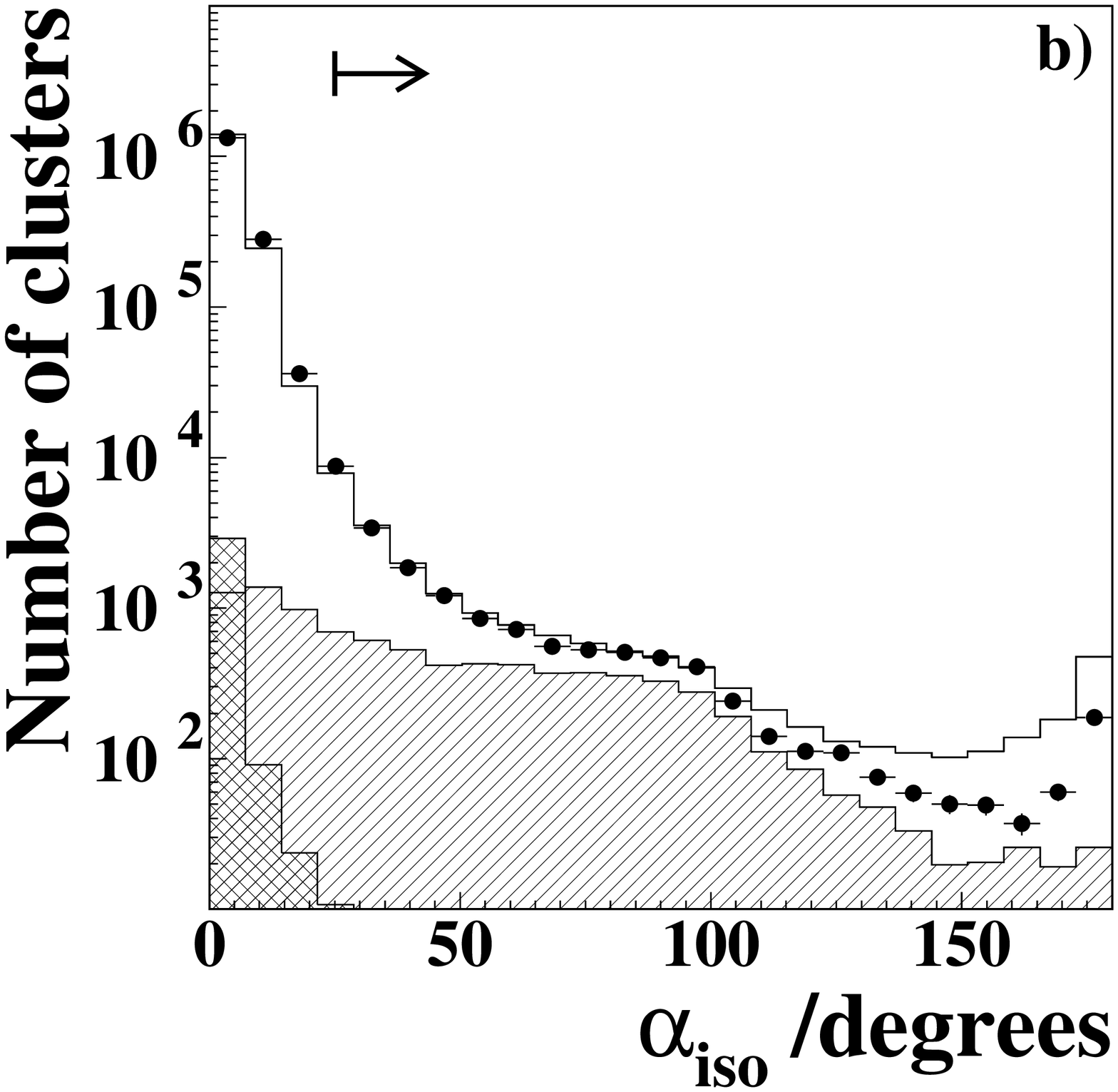} \\
\includegraphics[width=0.4\textwidth]{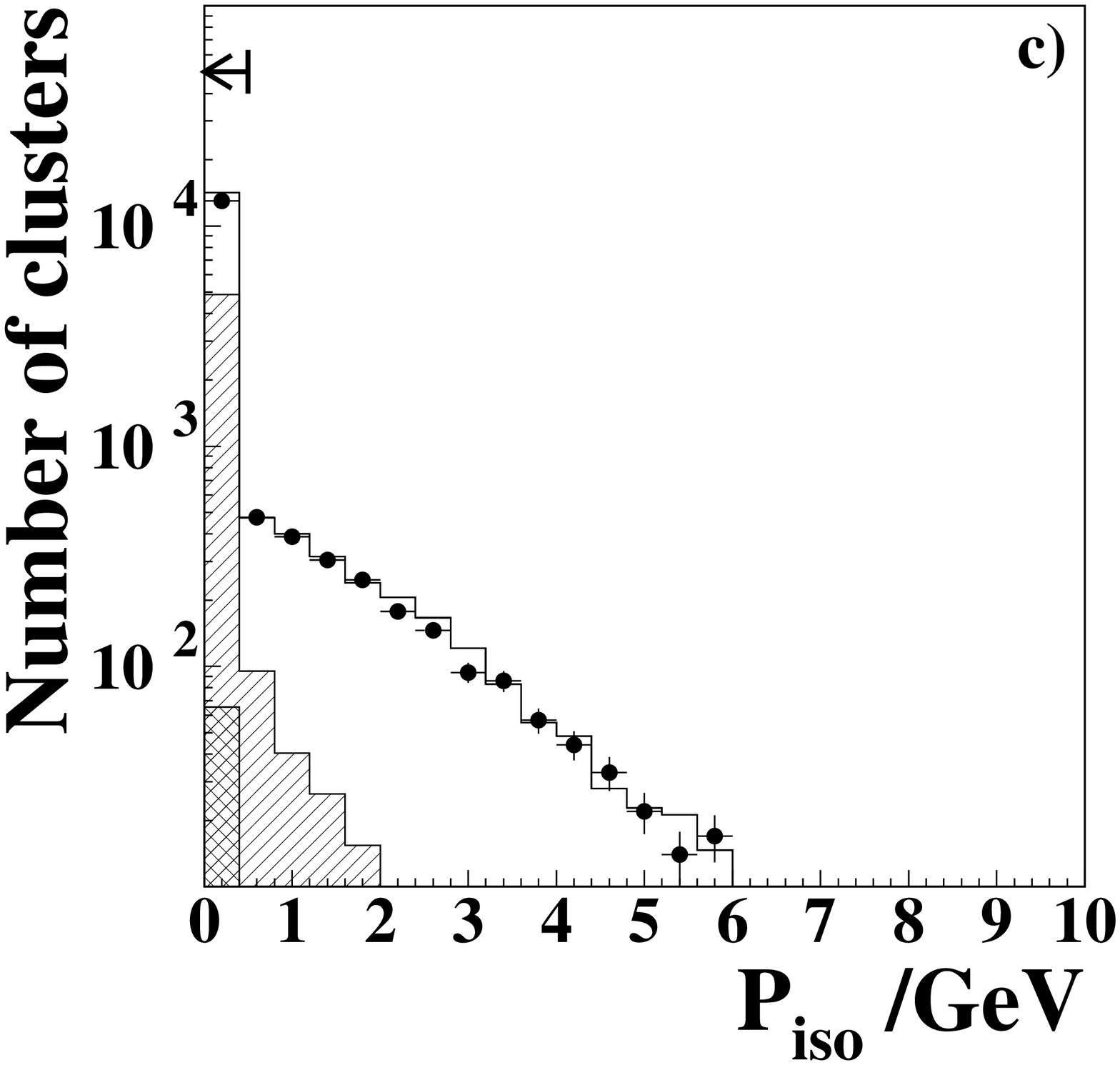} &
\includegraphics[width=0.4\textwidth]{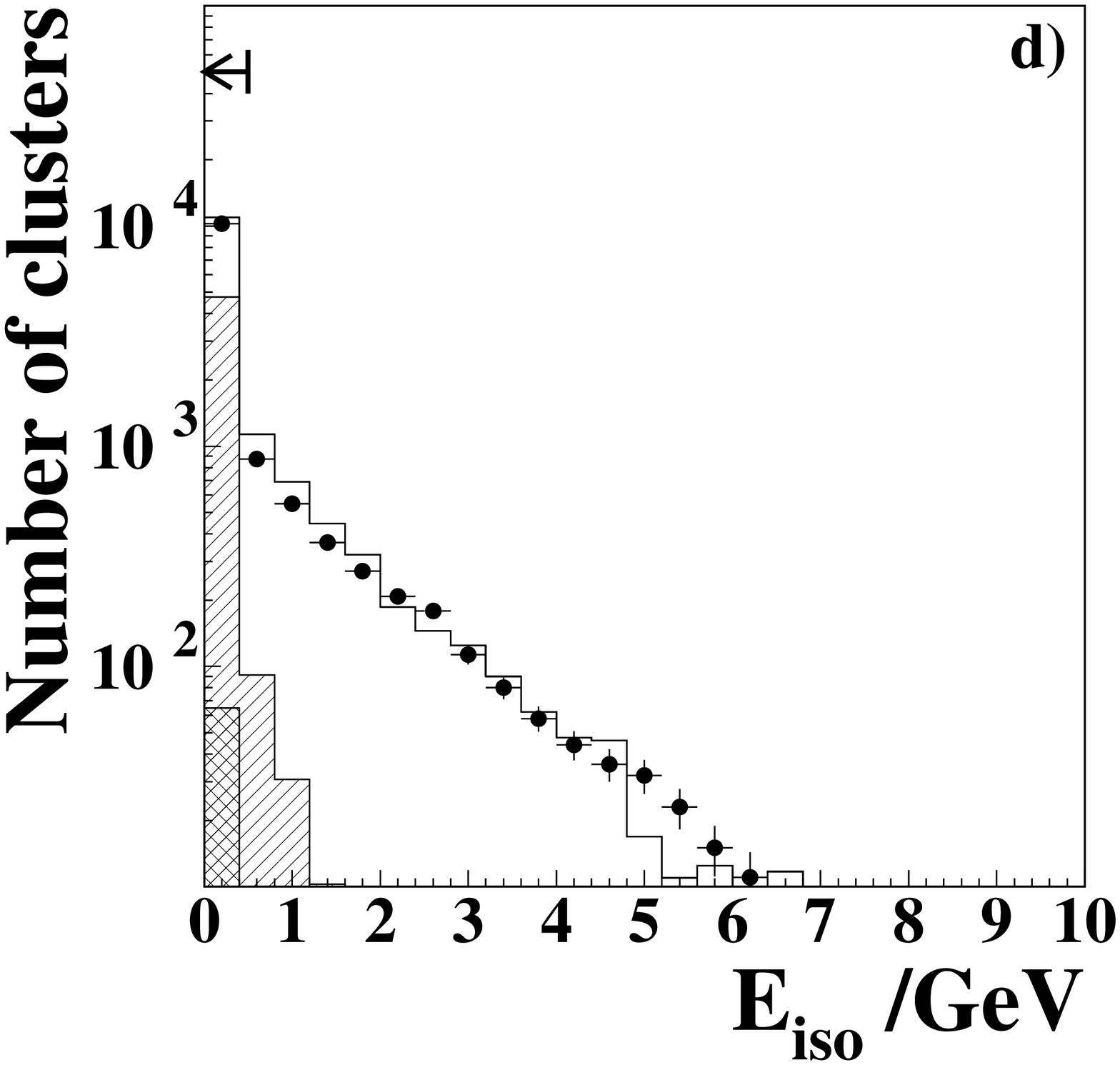} \\
\includegraphics[width=0.4\textwidth]{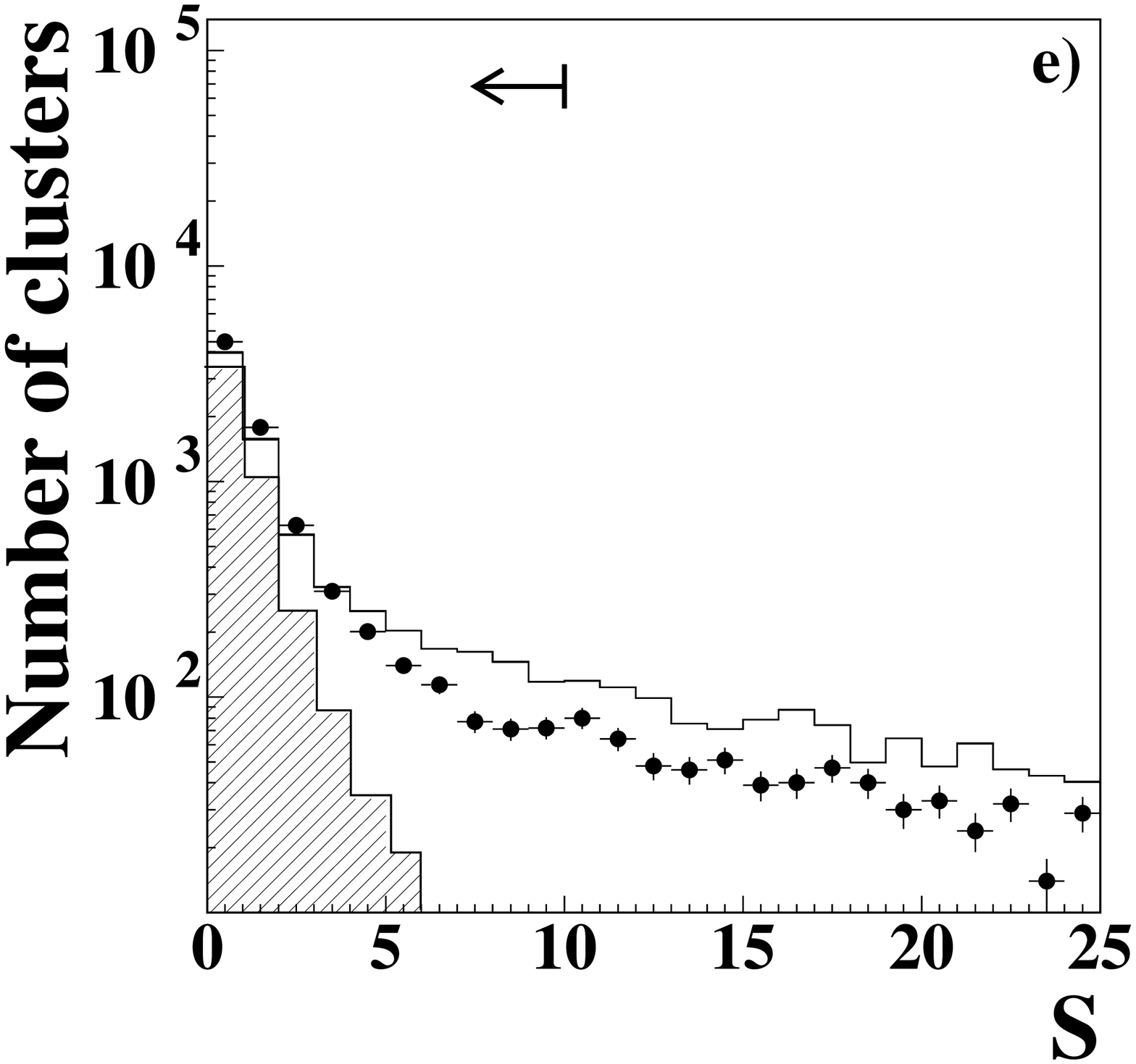} &
\includegraphics[width=0.4\textwidth]{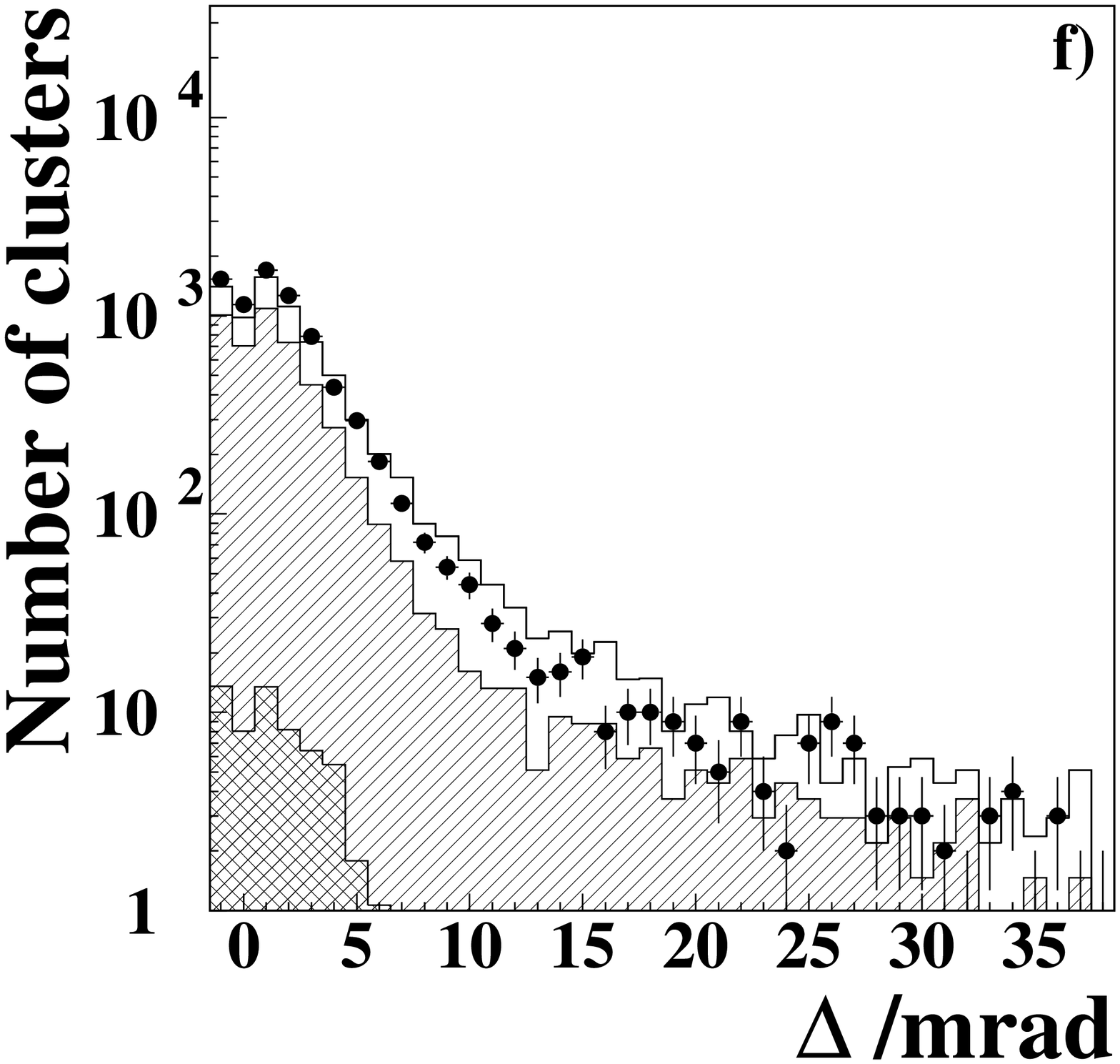} \\
\end{tabular}
\caption{Distributions of each variable used in the isolated photon
  selection. The error bars show the statistical errors. Monte Carlo
  distributions are normalized to the integrated luminosity of the
  data and the cross section of the process. Arrows in the figures
  show the selected region.  Distributions for radiative
  multi-hadronic events, which are signal events in this analysis, are
  overlaid on distributions for all multi-hadronic events and
  $\tau\tau$ events.  The distribution of each variable is obtained with
  the cuts on the preceeding variables applied.}
\label{SelectionVariables}
\end{figure}

Isolated photons are selected in these hadronic events as follows.
Electromagnetic clusters with an energy $E_{\mathrm{EC}}>10$~GeV are
chosen in order to suppress background from soft photons coming from
the decay of mesons.  Accordingly, our signal event is defined as an
$\epem\rightarrow\qqbar$ event with an ISR or FSR photon with energy
greater than 10~GeV.  We use electromagnetic clusters in the polar
angle region $|\cos{\theta_{\mathrm{EC}}}|<0.72$ corresponding to the
barrel of the detector, where there is the least material in front of
the lead glass, see Figure~\ref{SelectionVariables}~a).  Also, the
non-pointing geometry of the endcap electromagnetic calorimeter
complicates the cluster shape fitting explained below.  The number of
clusters in the data which satisfy $E_{\mathrm{EC}}>10$~GeV and
$|\cos{\theta_{\mathrm{EC}}}|<0.72$ is 1,797,532.  According to the
Monte Carlo simulation, 99.3\% of these selected clusters come from
non-radiative multi-hadronic events.

The candidate clusters are required to be isolated from
any jets, and from other clusters and tracks:
\begin{itemize}

\item The angle with respect to the axis of any jet, \Aiso, is required to
  be larger than $25^\circ$, see Figure~\ref{SelectionVariables}~b).  The
  jets are reconstructed from tracks and electromagnetic clusters,
  excluding the candidate cluster, using the Durham
  algorithm\cite{Catani:1991hj} with $y_{\mathrm{cut}}=0.005$.

\item The sum of the momenta \Piso\ of 
  tracks falling on the calorimeter surface inside a 0.2 radian cone
  around the photon candidate is required to be smaller than
  0.5~$\mathrm{GeV}/c$ (Figure~\ref{SelectionVariables}~c)).  The total
  energy deposition in the electromagnetic calorimeter within a cone
  of 0.2 radian around the photon candidate, \Eiso, is also required
  to be less than 0.5~GeV (Figure~\ref{SelectionVariables}~d)). 

\end{itemize}
After the isolation cuts, 11,265 clusters are retained.  The fraction
of clusters from non-radiative multi-hadronic events is reduced to
52.8\%.  The background from $\tau^+\tau^-$ events (two-photon events)
is 0.6\% (0.01\%)~\cite{toyaphd}.

\subsubsection{Likelihood Photon Selection}
\label{sec_llphotonselection}

Isolated photon candidates are selected by using a likelihood ratio
method with four input variables, see appendix~\ref{app_like} for
details.  The first two variables are $|\cos{\theta_{\mathrm{EC}}}|$
and \Aiso, defined above.  Two more variables, the cluster shape fit
variable $S$ and the distance $\Delta$ between the electromagnetic
calorimeter cluster and the associated presampler cluster, defined as
follows, reduce the background from clusters arising from the decays
of neutral hadrons.

The cluster shape fit variable, $S$, is defined by
\begin{equation}
  S = \frac{1}{N_{\mathrm{block}}}\sum_i{ 
\frac{(E_{\mathrm{meas},i}-E_{\mathrm{exp},i})^2}{\sigma_{\mathrm{meas},i}^2}}
\label{EqnDefCVariable}
\end{equation}
where $N_{\mathrm{block}}$ is the number of lead glass blocks included
in the electromagnetic cluster, $E_{\mathrm{meas},i}$ is the measured
energy deposit in the $i$th block, $E_{\mathrm{exp},i}$ is the
expected energy deposit in the $i$th block, assuming that the energy
is deposited by a single photon, and $\sigma_{\mathrm{meas},i}$ is the
uncertainty in the energy measured by $i$th electromagnetic
calorimeter block.  $E_{\mathrm{exp},i}$ is a function of position and
energy of the incident photon based on the simulation of the OPAL
detector with single photons.  The value of $S$ is determined by
minimizing Equation~(\ref{EqnDefCVariable}) under variation of the
position and energy of the cluster.  For a cluster to be considered
further in the likelihood, preselection cuts are applied: we require
the number of blocks to be at least two and the value of $S$ after the
fit to be smaller than 10.  The quality of the cluster shape fits
depends on the assumed resolution $\sigma_{\mathrm{meas},i}$; this
will be studied as a systematic uncertainty.

The variable $\Delta$ measures the distance between the
electromagnetic calorimeter cluster and the associated presampler
cluster, $\Delta=\max(|\Delta\phi|,|\Delta\theta|)$, with $\Delta\phi$
and $\Delta\theta$ the angular separations between the clusters.

The distributions of $S$ and $\Delta$ are shown in
Figures~\ref{SelectionVariables}~e) and~f).  The Monte Carlo
distributions in these figures are normalized according to the
luminosity obtained from small angle Bhabha events.

A disagreement between data and Monte Carlo is seen for $S$ and
\Aiso.  The level of agreement between data and Monte Carlo for the $S$
distribution is studied with photons in radiative muon pair events and
$\pi^0$s produced in tau pair events.  It is confirmed that the Monte
Carlo adequately reproduces the $S$ distributions~\cite{toyaphd}.  The
disagreement between data and Monte Carlo for distributions of $S$ and
\Aiso\ stems from the failure of the Monte Carlo generators to
correctly predict the rate of isolated neutral hadrons, as explained
in Section~\ref{BkgEstimation}.  In this analysis, the rate of
isolated neutral hadrons used in the background subtraction is
estimated from data by methods described in
Section~\ref{BkgEstimation}.

\begin{figure}[p]
\opalh\vspace{-0.3cm}
\centering
\includegraphics[width=\textwidth]{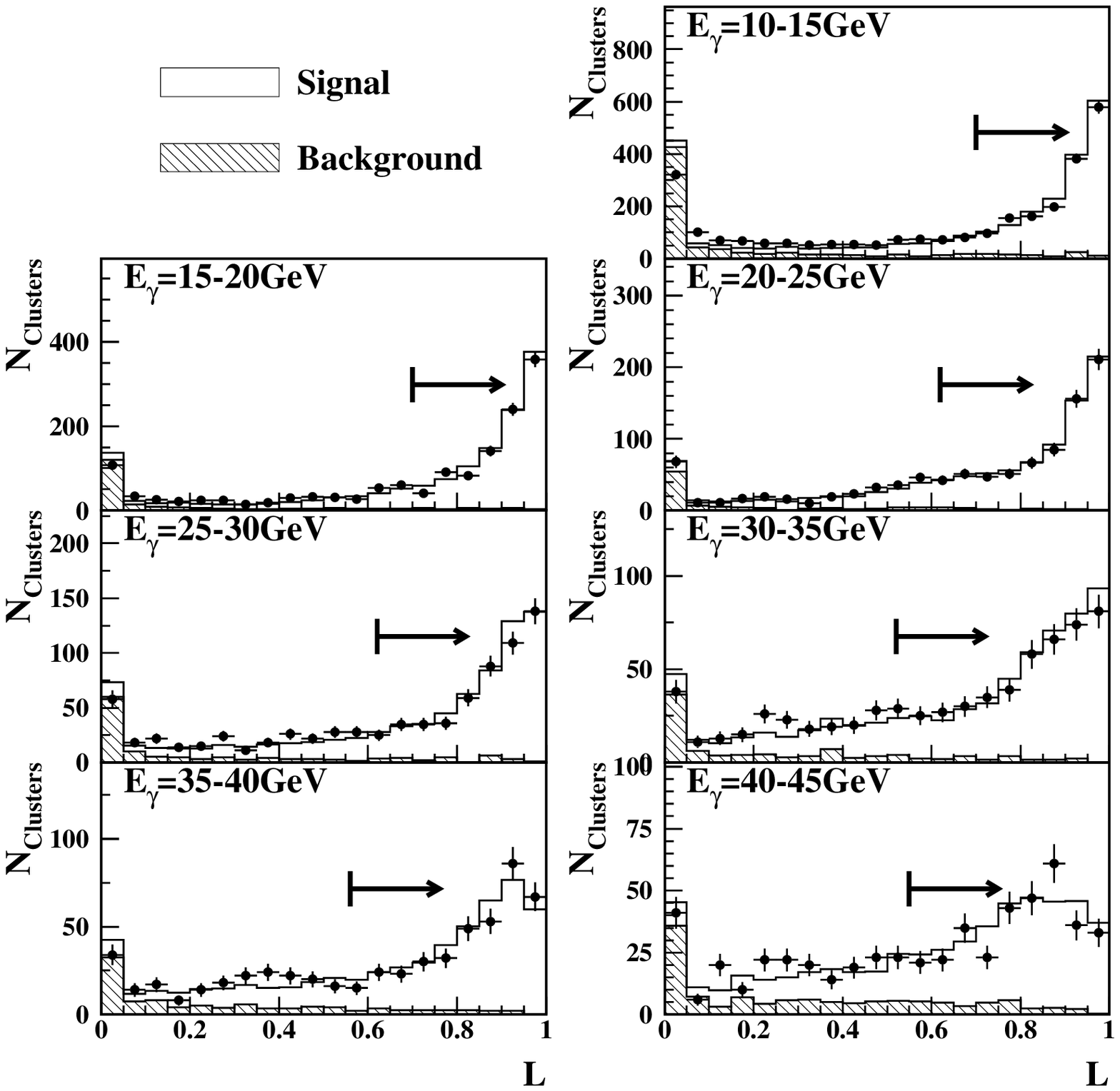}
\caption{Photon likelihood distributions. The error bar shows
  statistical error.  The Monte Carlo distributions are normalised to
  the total number of candidates in the data, and the neutral hadron
  background fractions are obtained from the fits described in
  Section~\ref{BkgEstimation}. The arrows indicate the selected
  regions.}
\label{LikelihoodDist}
\end{figure}

The likelihood calculation is performed with reference histograms made
for seven subsamples, chosen according to the cluster energy.  The cut
on the likelihood value is chosen so as to retain 80\% of the signal
events.  The likelihood distributions for data and Monte Carlo are
shown in Figure~\ref{LikelihoodDist}.  It can be seen that the
likelihood distributions for signal and background events are well
separated for each region of electromagnetic cluster energy.
Electromagnetic clusters which pass the likelihood selection are
regarded as photon candidates. If more than one candidate is found in
the same event the one with the highest energy is chosen.

\subsection{Final Data Sample}

Hadronic events with hard isolated photon candidates are divided into
seven subsamples according to the photon energy for further analysis.
Table~\ref{SelectedEvents} shows the mean values of \rsp, the number
of data events and the number of background events for each subsample.

\subsection{Background Estimation}
\label{BkgEstimation}

According to the Monte Carlo simulation, the contamination from $\tau$
pair events is between 0.5 and 1.0\%. The impact of this small number
of events is further reduced because the value of event shape
observables for $\tau$ pair events are concentrated in the lowest bin
of the distributions, outside the fitting range, so their effect on
the \as\ fits is negligible.  The contribution of two photon processes
is less than 0.01\% in all subsamples and is ignored.

As mentioned in~\cite{PromptPhoton}, the \Jetset\ Monte Carlo fails to
reproduce the observed rate of isolated electromagnetic clusters, both
for isolated photons and isolated $\pi^0$'s.  Isolated neutral hadrons
are the dominant source of background for this analysis, and their
rate has been estimated from data using the following two methods.

Firstly, with the likelihood ratio method the observed likelihood
distributions in the data in bins of photon energy were fitted with a
linear combination of the Monte Carlo distributions for signal and
background events which pass the isolation cuts and likelihood
preselection requirements.  The overall normalisation of the Monte
Carlo distribution is fixed to the number of data events.  The fit
uses a binned maximum likelihood method with only the fraction of
background events as a free parameter.  Figure~\ref{LikelihoodDist}
shows the fit results.  The values of \chisqd\ are
between 1.2 and 3.4 for 18 degrees of freedom.

Secondly, with the isolated tracks method the fraction of background
from isolated neutral hadrons was estimated from the rates of isolated
charged hadrons.  We select from the data tracks which satisfy the
same isolation criteria as the photon candidates. The composition of
these isolated charged hadrons obtained from \Jetset\ is used to infer
the rates of charged pions, kaons and protons.  When isospin symmetry
is assumed, the rates of neutral pions, neutral kaons and neutrons can
be estimated from the rates of charged pions, charged kaons and
protons, respectively:
\begin{equation}
  \label{IsospinSymmetry}
  R_{\pi^0}=\frac{1}{2}R_{\pi^{\pm}}\hspace{1cm} 
  R_{\mathrm{K}^0_{\mathrm{L}}}=\frac{1}{2}R_{\mathrm{K}^{\pm}}\hspace{1cm}  
  R_{\mathrm{n}}=R_{\mathrm{p}}\ ,
\end{equation}
where $R_{\mathrm{X}}$ is the production rate of particle X.
According to \Jetset\ tuned with OPAL data, the rate of isospin
symmetry violation is 10\% for pions and 5\% for kaons and
protons.  This is assigned as a systematic uncertainty for the isolated
tracks method and combined with the statistical uncertainty.

The neutral hadron background fractions estimated by these two methods
are shown in Table~\ref{SelectedEvents}.  The statistical errors from
the number of data and Monte Carlo events from fitting the likelihood
distributions are shown.  The results from the two methods are within
at most three standard deviations of these errors, except in the
$E_{\gamma}$ bin $35-40$~GeV.

The standard analysis will use the likelihood ratio method.  Any
differences in the resulting values of $\as$ obtained by using the two
background estimate methods will be treated as a systematic
uncertainty.

%%%%%%%%%%%%%%%%%%%%%%%%%%
\section{Measurement of Event Shape Distributions}
\label{MeasureEventShape}
%%%%%%%%%%%%%%%%%%%%%%%%%%

In this analysis event shape observables as defined above in
section~\ref{sec_method} are calculated from tracks and
electromagnetic clusters excluding the isolated photon candidate.  The
contributions of electromagnetic clusters originating from charged
particles are removed by the method described in~\cite{CharginoMT}.

We evaluate the observables in the cms frame of the hadronic system.
The Lorentz boost is determined from the energy and angle of the
photon candidate.  When the four-momentum of particles in the hadronic
system is calculated, electromagnetic clusters are treated as photons
with zero mass while tracks of charged particles are treated as
hadrons with the charged pion mass.

\begin{figure}[htb]
\opalh\vspace{-0.2cm}
\centering
\includegraphics[width=\textwidth]{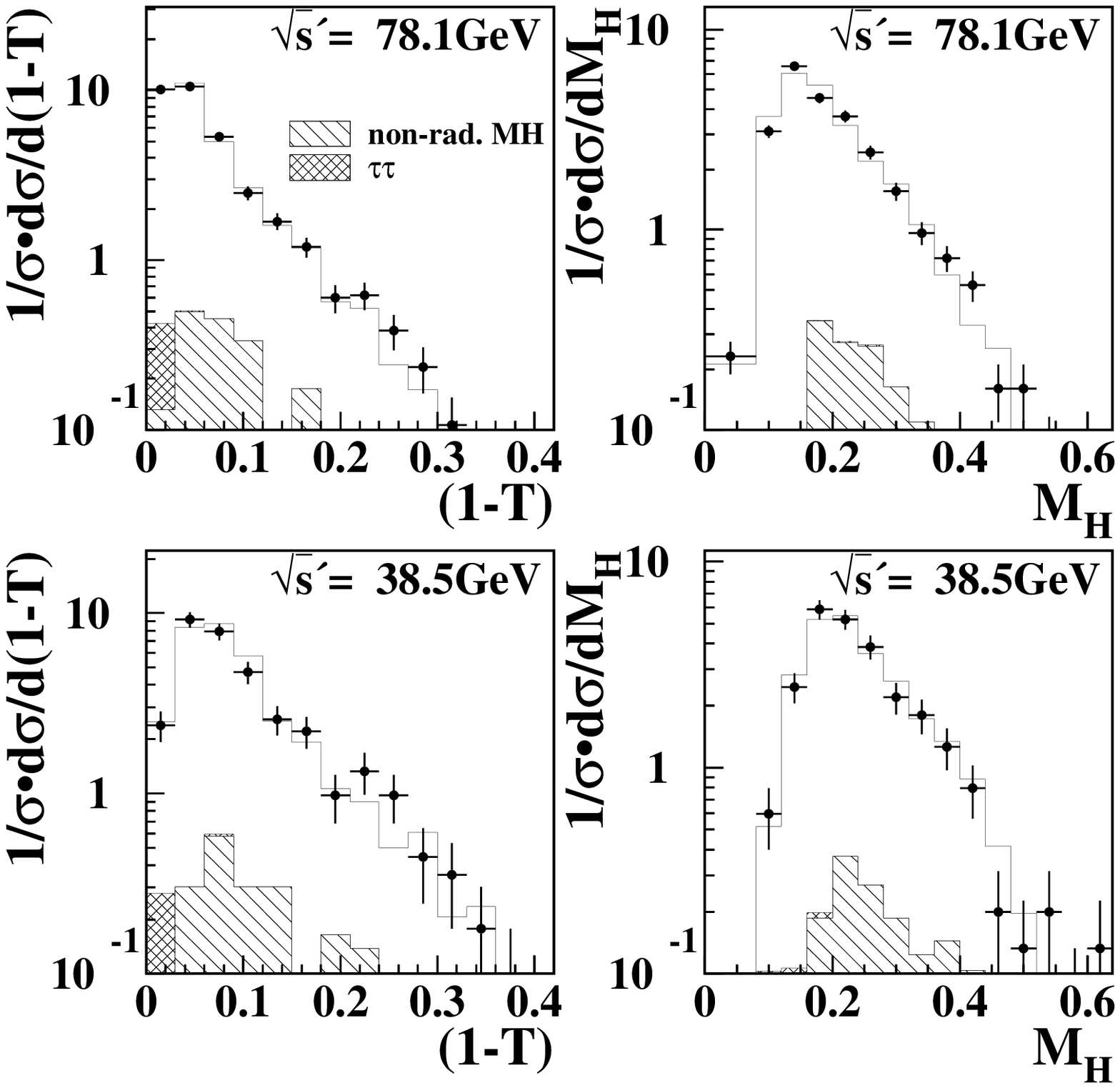} 
\caption{Event shape distributions before background subtraction and 
  detector correction. Two of the six event shape observables, \thr\
  and \mh, are shown for the low (38.5~GeV) and high (78.1~GeV) \rsp\
  samples. The histograms show Monte Carlo distributions.  The error
  bars show the statistical errors.}
\label{DetectorLevel}
\end{figure}

Distributions of the event shape observables (\thr) and \mh\ are shown
for two cms energies in Figure~\ref{DetectorLevel}.  The remaining
background is removed by subtracting the scaled Monte Carlo
predictions for non-radiative hadronic events and for $\tau$ pair
events using the background estimates listed in
Table~\ref{SelectedEvents}.  The effects of the experimental
resolution and acceptance are unfolded using Monte Carlo samples with
full detector simulation (detector correction).  The unfolding is
performed bin-by-bin with correction factors $r_i^{Det}=h_i/d_i$,
where $h_i$ represents the value in the $i$th bin of the event shape
distribution of stable hadrons in the Monte Carlo simulation, where
``hadrons'' are defined as particles with a mean proper lifetime
longer than $3\cdot 10^{-10}$s.  $d_i$ represents the value in the
$i$th bin of the event shape distribution calculated with clusters and
tracks obtained from Monte Carlo samples with detector simulation
after the complete event selection has been applied.  We refer to the
distributions after applying these corrections as data corrected to
the hadron level.

\begin{figure}[htb]
\opalh\vspace{-0.7cm}
\centering
\begin{tabular}{cc}
\includegraphics[width=0.45\textwidth]{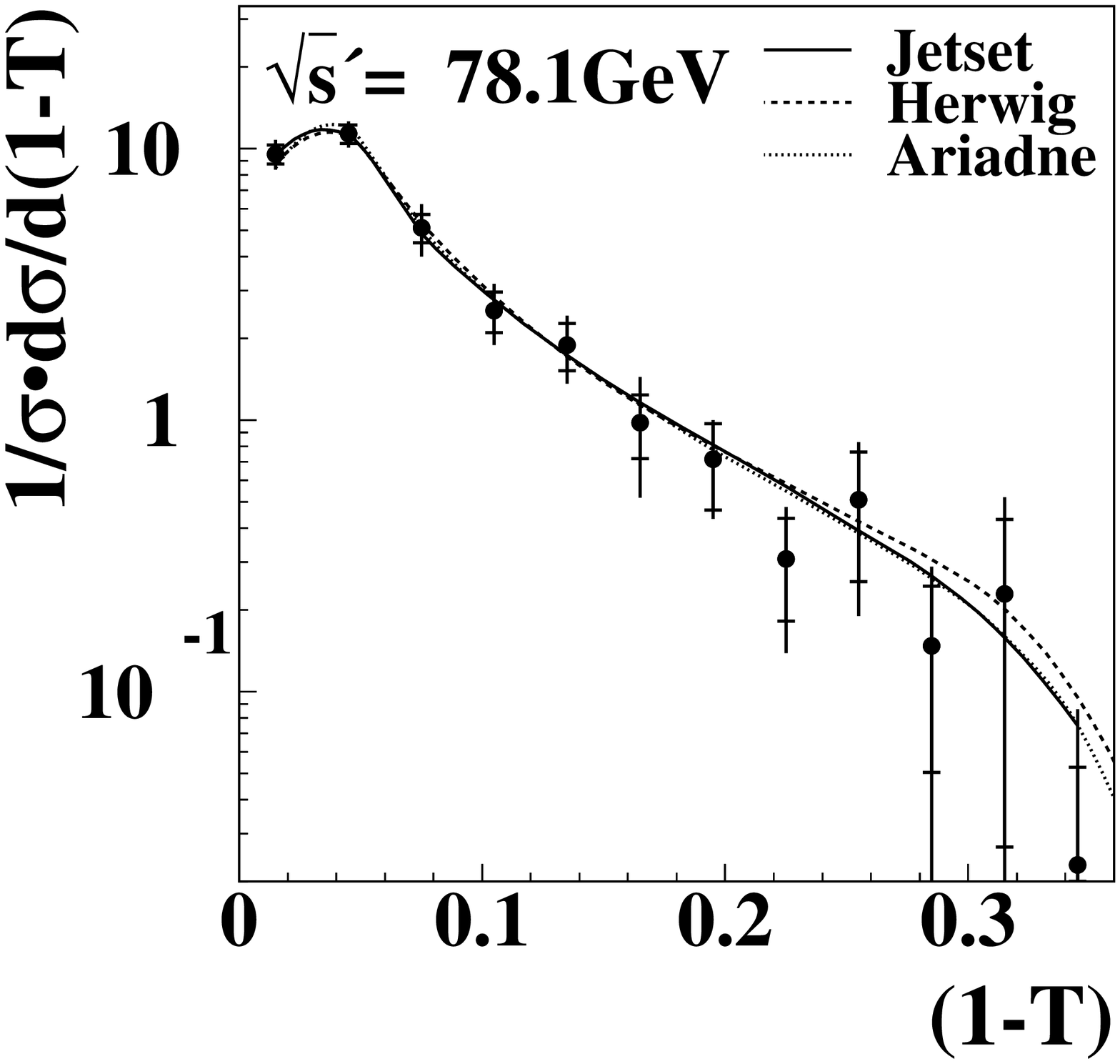} &
\includegraphics[width=0.45\textwidth]{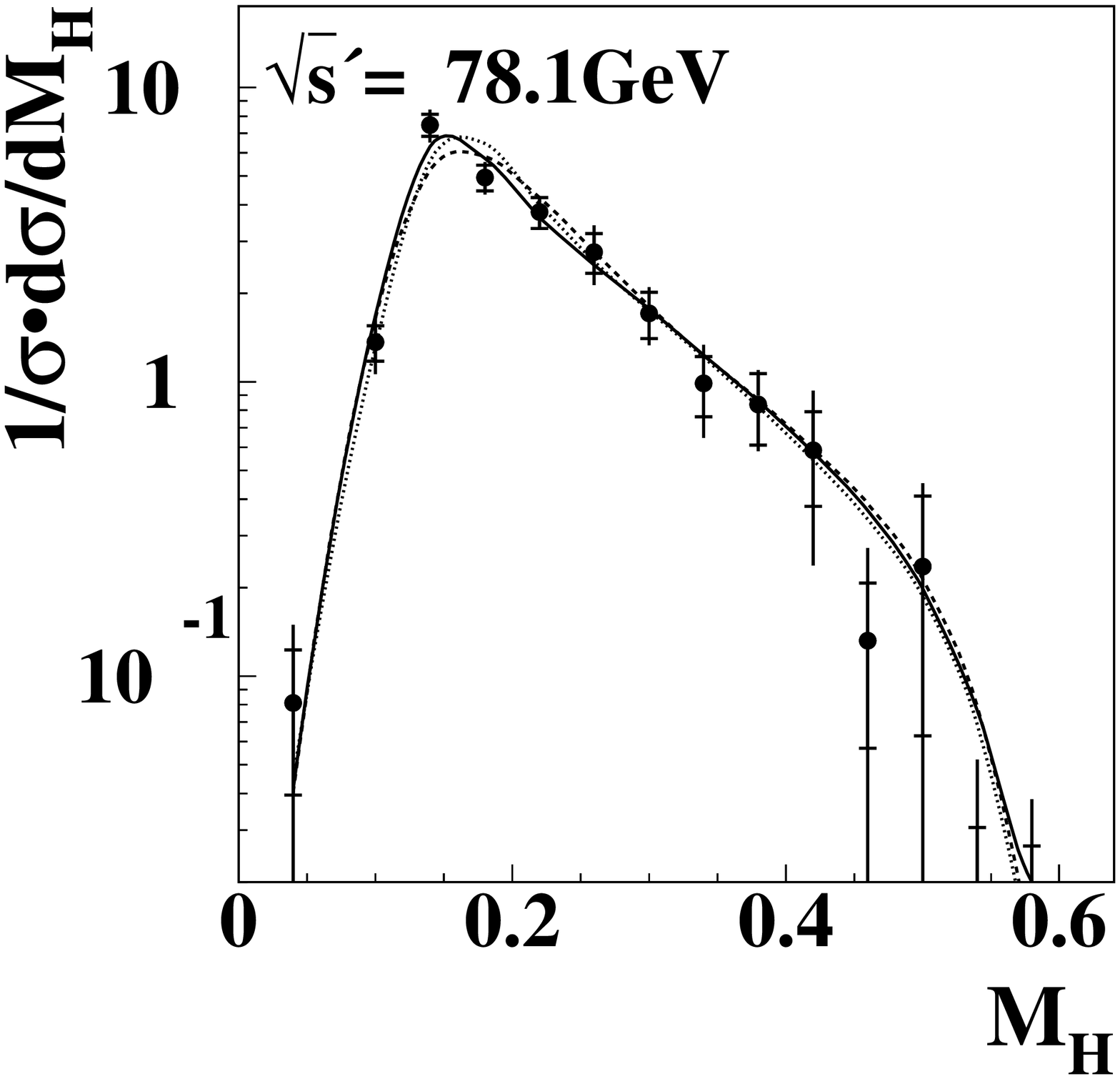} \\
\includegraphics[width=0.45\textwidth]{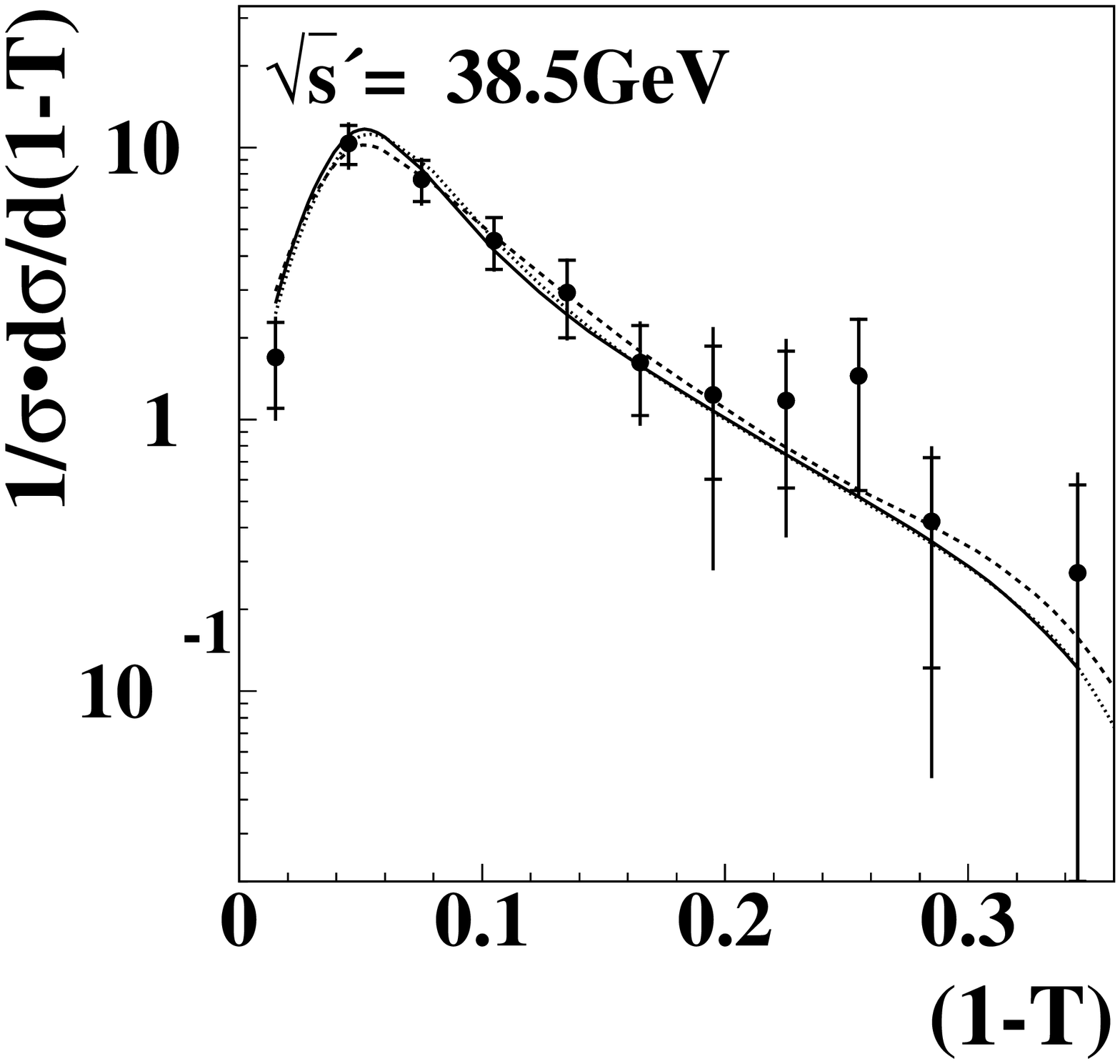} &
\includegraphics[width=0.45\textwidth]{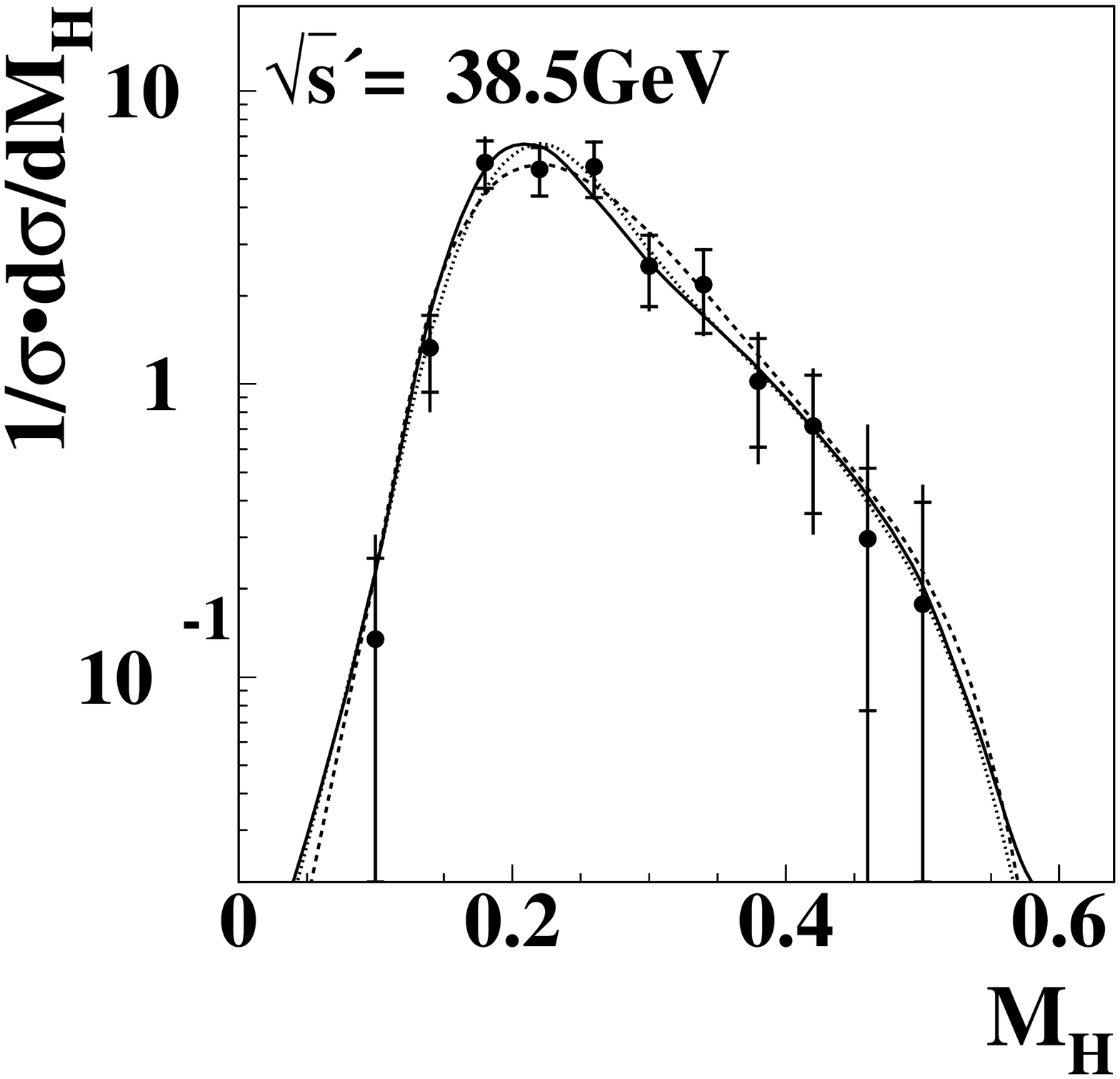} \\
\end{tabular}
\caption{Event shape distributions at the hadron level. The error bars
  correspond to the statistical and experimental uncertainties
  described in Section~\ref{SystUncertaintiesExpt}. Two of the six
  event shape observables, \thr\ and \mh, are shown for the low
  (38.5~GeV) and high (78.1~GeV) \rsp\ samples.  The small lines on
  the error bars show the extent of the statistical uncertainty.  The
  data points are placed at the centres of the corresponding bins.
  The predictions of \Jetset, \Herwig\ and \Ariadne\ at the
  corresponding \rsp\ values are also shown as lines.}
\label{HadronLevel}
\end{figure}

The distributions of the event shape observables \thr\ and \mh\ for
data corrected to the hadron level and corresponding Monte Carlo
predictions are shown in Figure~\ref{HadronLevel}\footnote{The values
of the six observables at the seven energy points are given
in~\cite{toyaphd} and will be available under {\tt
http://durpdg.dur.ac.uk/HEPDATA/}.}.  The Monte Carlo samples are
generated with cms energies set to the mean value of \rsp\ in each
subsample.  In the production of the Monte Carlo samples ISR and FSR
is switched off and on, respectively.  The predictions from the event
generators are consistent with the data for all \rsp\ bins.  There is
similar agreement between data and event generator predictions for the
other observables.

%%%%%%%%%%%%%%%%%%%%%%%%%%
\section{Measurement of \mathversion{bold}$\as$\mathversion{normal}}
\label{EventShapeAs}
%%%%%%%%%%%%%%%%%%%%%%%%%%

The measurement of \as\ is performed by fitting perturbative QCD
predictions to the event shape distributions corrected to the hadron
level for (\thr), \mh~\cite{CalcThrustMH}, \bt,
\bw~\cite{CalcBTBWOld,CalcBTBW}, $C$~\cite{Parisi:1978eg,Donoghue:1979vi} and
$y_{23}^D$~\cite{Catani:1991hj,Bethke:1991wk,Brown:1991hx}.  The
$\mathcal{O}(\as^2)$ and NLLA calculations are combined with the
ln($R$) matching scheme. The effects of hadronisation on event shapes
must be taken into account in order to perform fitting at the hadron
level (hadronisation correction).  Preserving the normalisation in the
hadronisation correction is not trivial for low \rsp\ samples because
of large hadronisation corrections.  The hadronisation correction is
applied to the integrated (cumulative) theoretical calculation to
conserve normalisation as in our previous
analyses~\cite{AsLEP2-130GeV,AsLEP2}.  The hadron level predictions
are obtained from the cumulative theoretical calculation multiplied by
a correction factor $R_i^{Had}=H_i/P_i$, where $P_i$ ($H_i$)
represents the value in the $i$th bin of the cumulative event shape
distribution calculated by Monte Carlo simulation without (with)
hadronisation.  The \Jetset\ Monte Carlo event generator is used for
our central results, while \Herwig\ and \Ariadne\ are considered as
alternatives for the estimation of systematic uncertainties.

The fit of the hadron level QCD predictions to the event shape
observables uses a least \chisq\ method with $\as(Q)$ treated as a
free parameter.  Only statistical uncertainties are taken into account
in the calculation of \chisq.  When the total number of events is
small, the differences between the statistical errors counting larger
or smaller numbers of events than the theoretical prediction can bias
the fit result.  In order to avoid this bias the value of the fitted
theoretical distribution is used to calculate the statistical error
instead of the number of events in each bin of the data distribution.
The statistical uncertainty is estimated from the fit results derived
from 100 Monte Carlo subsamples with the same number of events as
selected data events.

\begin{figure}[htb]
\vspace*{-1.0cm}
\opalh\vspace{-0.2cm}
\centering
\begin{tabular}{cc}
\includegraphics[width=0.4\textwidth]{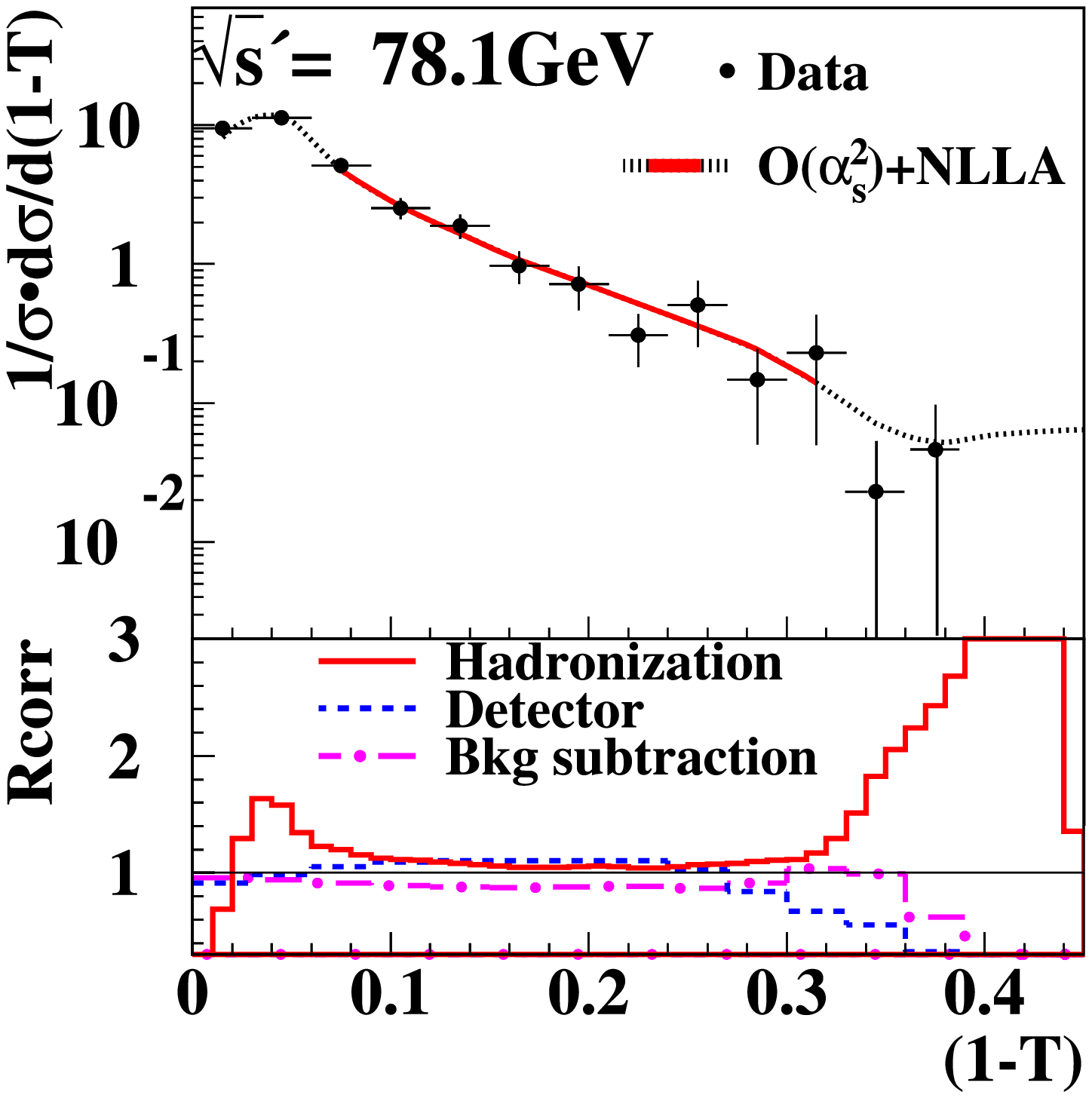} &
\includegraphics[width=0.4\textwidth]{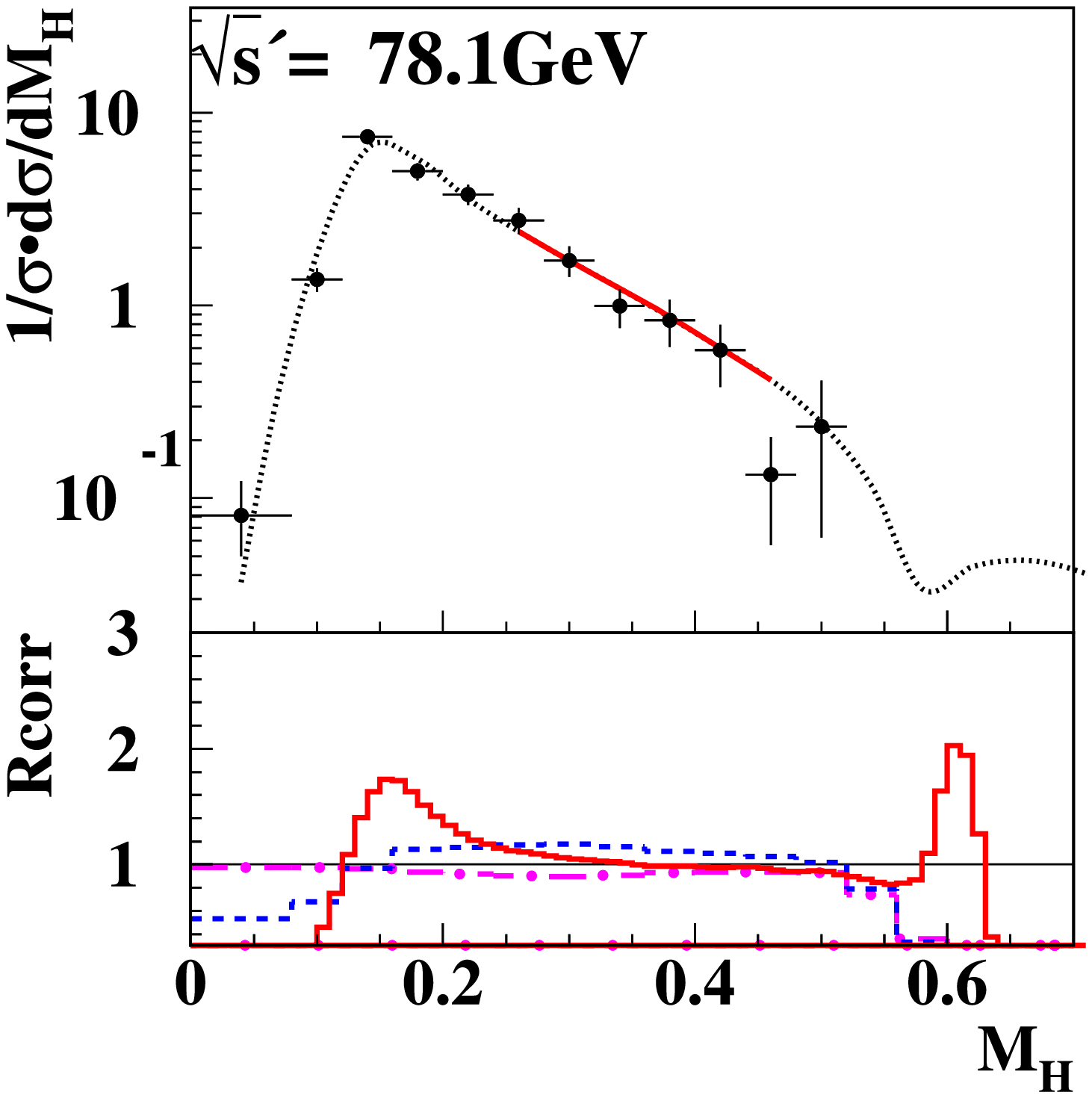} \\
\includegraphics[width=0.4\textwidth]{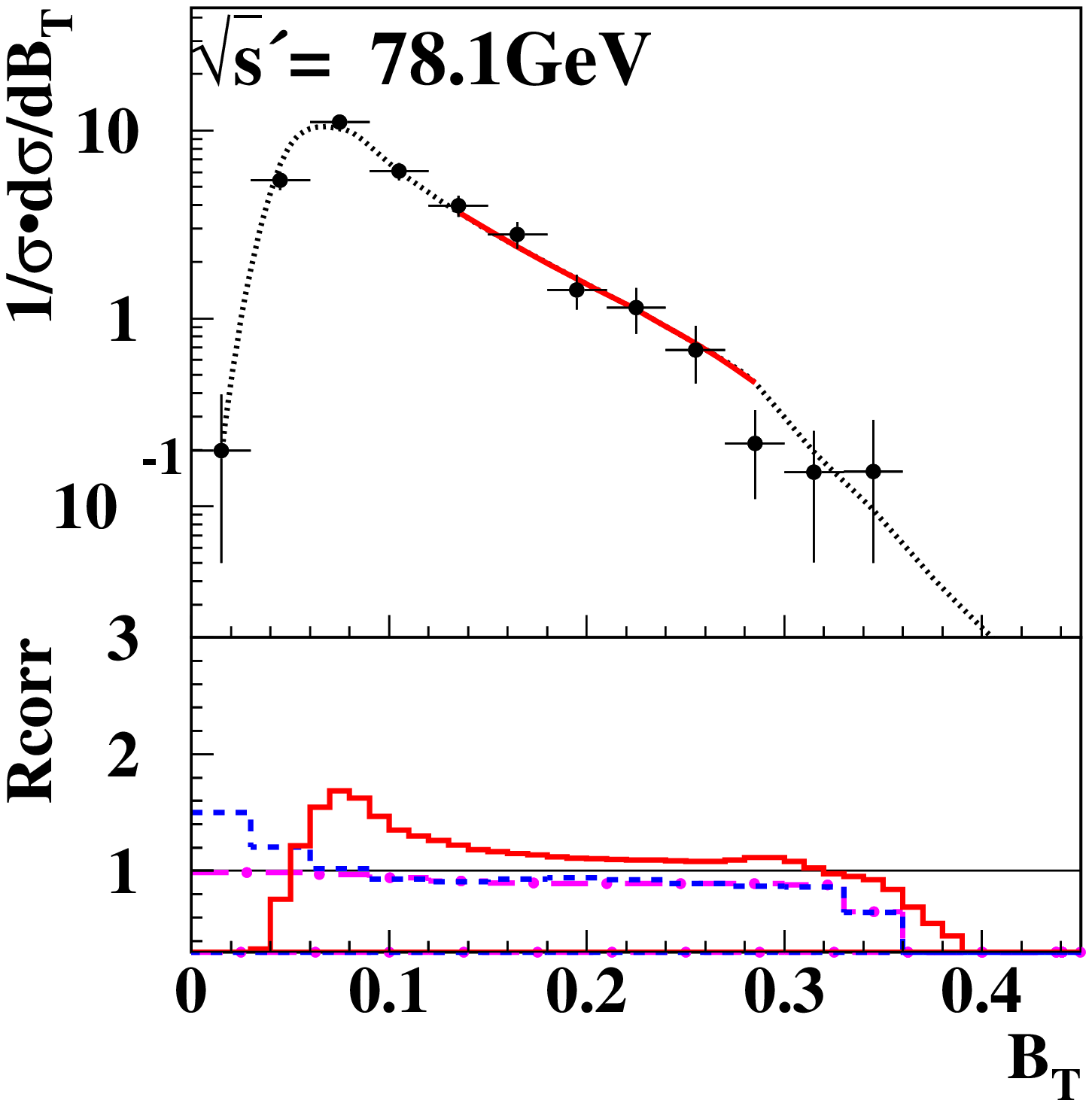} &
\includegraphics[width=0.4\textwidth]{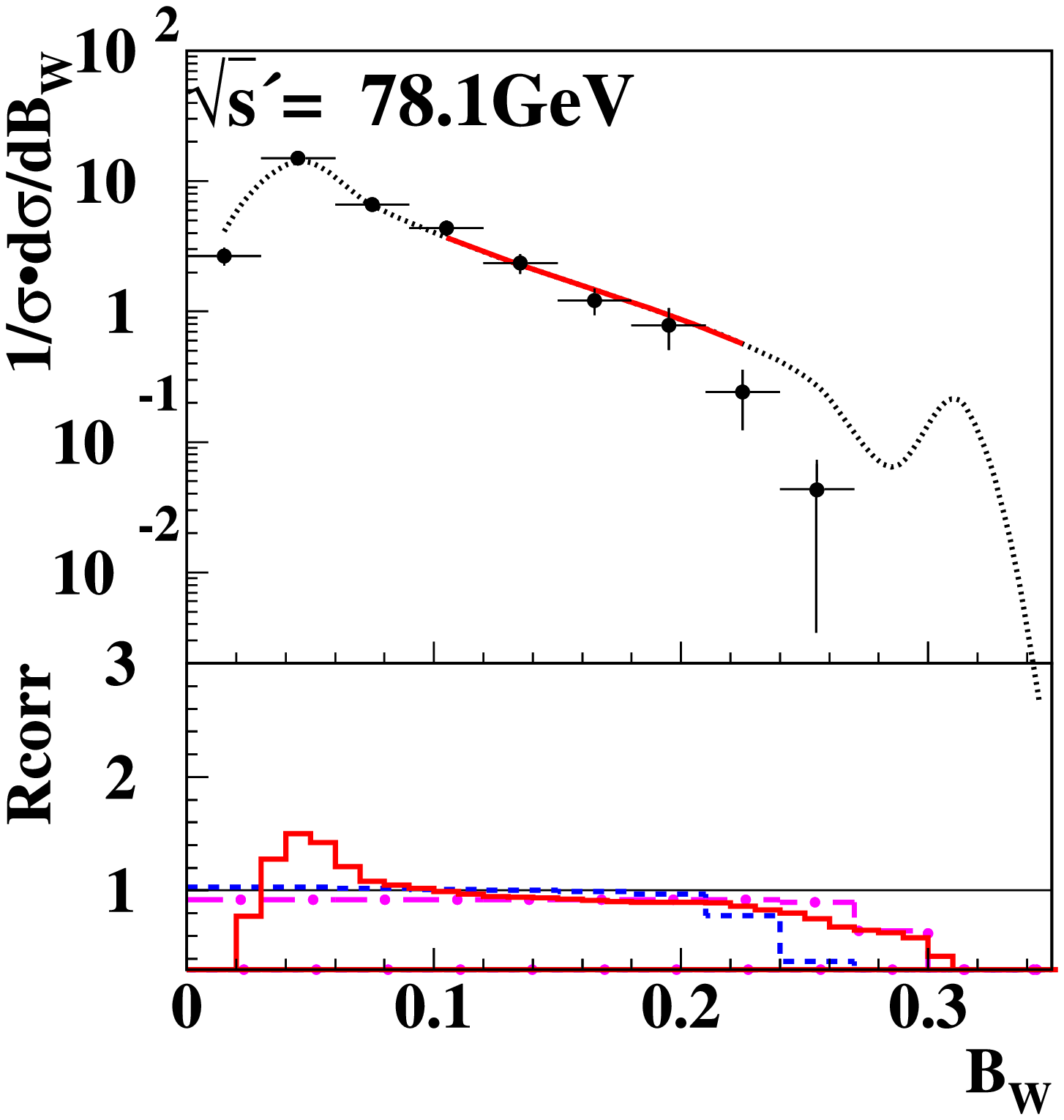} \\
\includegraphics[width=0.4\textwidth]{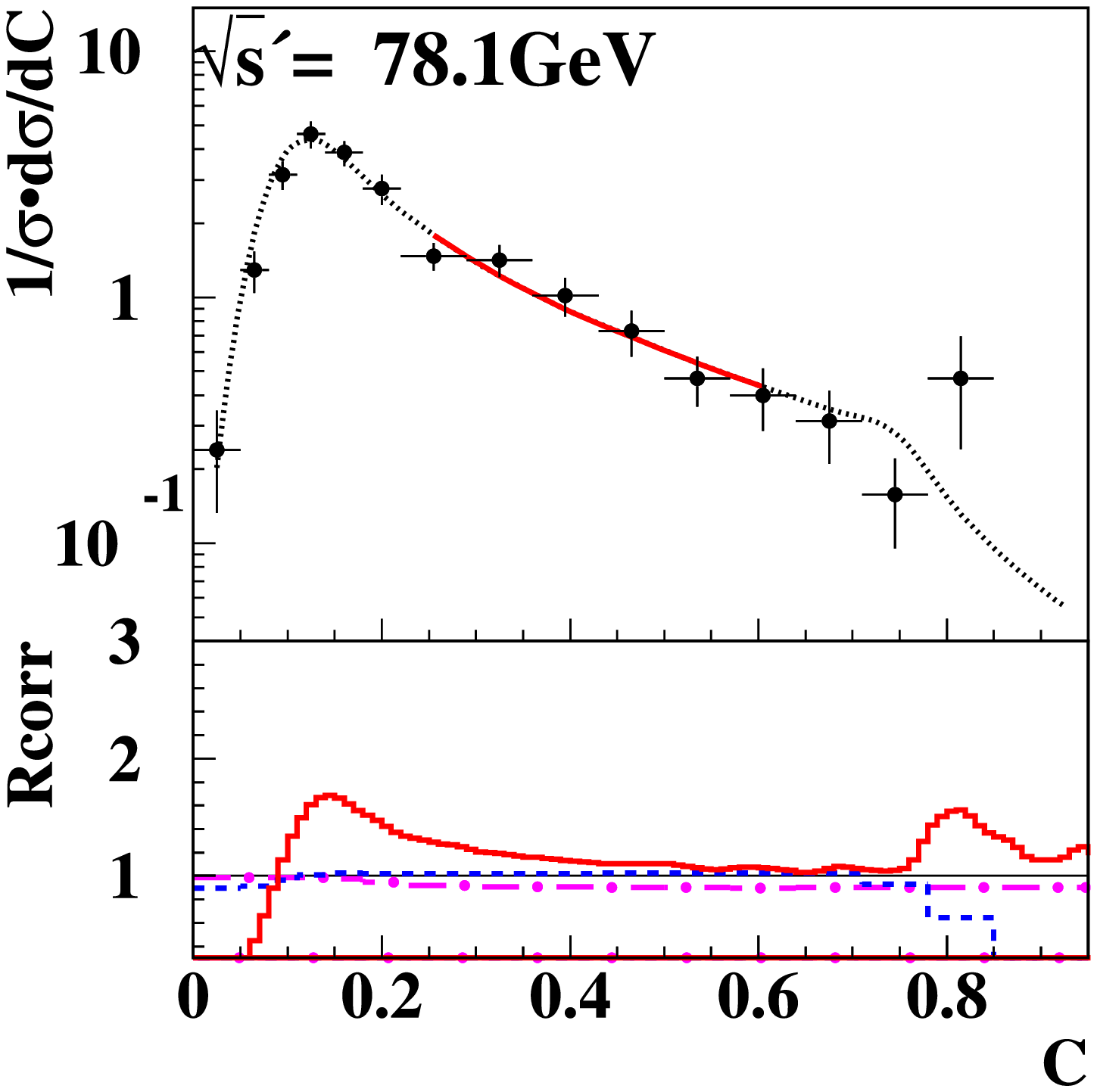} &
\includegraphics[width=0.4\textwidth]{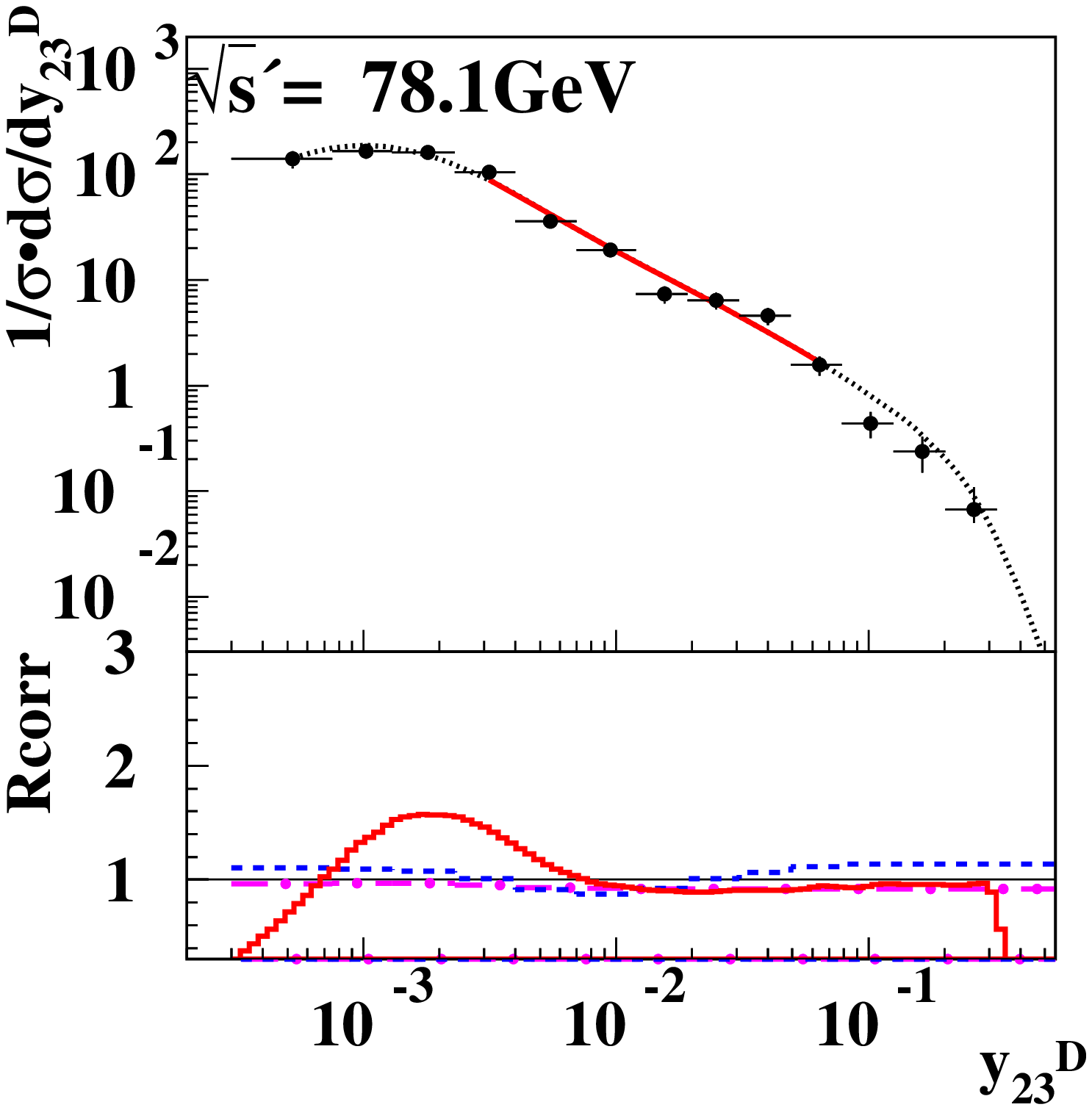} \\
\end{tabular}
\caption{Event shape distributions for data at $\langle \sqrt{s'} 
  \rangle = 78.1$~GeV and the fitted theoretical predictions.  The
  error bars show the statistical errors.  The solid lines in the
  theoretical predictions show the regions used in the fit. Three
  corrections are plotted as ``$R\mathrm{corr}$'': the detector
  correction, $r_i^{Det}$ (dashed line), the hadronisation correction,
  $R_i^{Had}$ (solid line), and the ratio of distributions after and
  before background subtraction (dotted line). The hadronisation
  correction is shown by the ratio of differential distributions in
  these figures (see text for details).}
\label{FittedDist}
\end{figure}

The region used in the fit is adjusted such that the background
subtraction and the detector and hadronisation corrections are small
(less than 50\%) and uniform in that region.  The resulting fit ranges
are mainly restricted by the hadronisation corrections.  The QCD
predictions at $\rsp=78$~GeV fitted to data after applying the
hadronisation correction are shown in Figure~\ref{FittedDist}.  Good
agreement between data and theory is seen.  The fitted values of \as\
and their errors for each event shape observable are shown in
Tables~\ref{FitResult-1}-\ref{FitResult-4}.

\subsection{Systematic Uncertainties}
\label{SystUncertainties}

\subsubsection{Experimental Uncertainties}
\label{SystUncertaintiesExpt}

The experimental uncertainty is estimated by adding in quadrature the
following contributions:
\begin{itemize}
\item the difference between the standard result and the result when all
  clusters and tracks are used without correcting for double counting
  of energy.  This variation is sensitive to imperfections of the
  detector simulation.
\item the largest deviation between the standard result and the result
  when the analysis is repeated with tighter selection criteria to
  eliminate background (standard values in brackets): the thrust axis
  is required to lie in the range $|\cos{\theta_T}|<0.7$ (0.9), or the
  cluster shape variable is required to be smaller than 5 (10), or the
  isolation angle from any jet is required to be larger than
  $35^\circ$ ($25^\circ$).
\item the difference between the standard value and the value 
  obtained by repeating the analysis with the background fractions
  estimated from the rate of isolated charged hadrons as described in
  Section~\ref{EventSelection}.
\item the difference between the standard result and the result when
  the single block energy resolution is varied to give the expected
  \chisq\ in the cluster shape fits.  This check is made, because the
  values of \chisq\ in the cluster shape fits depend on the assumed
  energy resolution.
\item the maximum difference between the standard result and the result when 
  the fit regions are varied.  The lower and upper limit of the 
  fitting region are independently changed by $\pm1$ bin. 
\end{itemize}
The tighter selection on $|\cos{\theta_T}|$ and the alternative single
block energy resolution of the electromagnetic calorimeter yield the
largest contributions to the experimental systematic uncertainty.  The
overall resolution and energy scale uncertainty of the electromagnetic
calorimeter have a neglegible effect on the results of this analysis.

\subsubsection{Hadronisation Uncertainties}

The following variations are performed in order to estimate
the hadronisation uncertainties:
\begin{itemize}
\item the largest of the changes in \as\ observed when independently 
  varying the hadronisation parameters $b$ and $\sigma_Q$ by $\pm 1$
  standard deviation about their tuned values in
  \Jetset~\cite{TuneJetset};
\item the change observed when the parton virtuality cut-off 
  parameter $Q_0$ is varied by $\pm 1$ standard deviation about
  its tuned value in \Jetset;
\item the change observed when only the light quarks u, d, s
  and c are considered at the parton level in order to estimate
  potential quark mass effects; 
\item the differences with respect to the standard result 
  when \Herwig\ or \Ariadne\ are used for the hadronisation
  correction, rather than \Jetset.
\end{itemize}
We define the hadronisation correction uncertainty by adding in
quadrature the deviation when using only light quarks and the larger
deviation when using \Herwig\ or \Ariadne\ to calculate the
corrections.  These variations are observed to lead to larger
differences than all other variations, i.e.\ the main contributions to
the hadronisation uncertainties are the choice of hadronisation model
and the potential effect of quark masses.

\subsubsection{Theoretical Uncertainties}

We fix the renormalisation scale parameter $\xmu\equiv\mu/Q$ to 1,
where $\mu$ is the energy scale at which the theory is renormalized
and $Q$ is the energy scale of the reaction.  Although the uncertainty
on the choice of the value of \xmu\ gives a large contribution to the
systematic uncertainty, the means of quantifying this uncertainty is
essentially arbitrary.  We define the scale uncertainty as the larger
of the deviations of \as\ when \xmu\ is changed from 1 to 0.5 or 2.0.

The {\cal O($\as^2$)} and NLLA calculations are combined with the
ln(R) matching scheme.  The variation in \assp\ due to using
different matching schemes is much smaller than the renormalisation
scale uncertainty~\cite{AsLEP1}, and is not included as an additional
theoretical systematic uncertainty.

\subsection{Combination of \mathversion{bold}$\as$\mathversion{normal} Results}
\label{Combination}

\begin{figure}[htb]
\centering
\vspace{-0.8cm}
\includegraphics[width=0.70\textwidth]{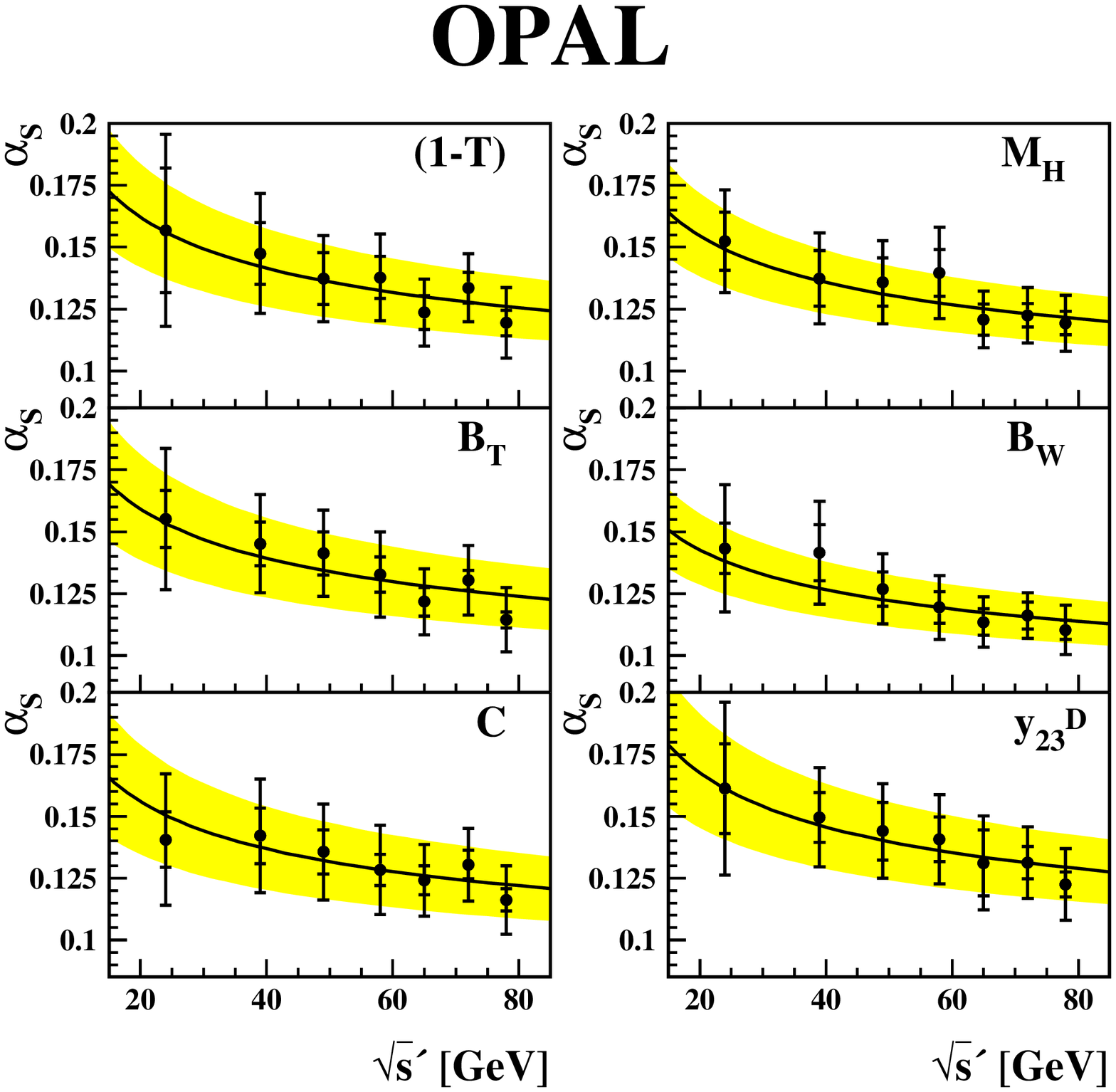}
\caption{Energy dependence of \as\ for all \rsp\ subsamples.  The
  inner error bars show the statistical and the outer error bars the
  total uncertainties.  The curves and shaded bands show the QCD
  prediction for the running of \as\ obtained with the corresponding
  values of \asmz\ with total errors from Table~\ref{CombinedResult}.}
\label{AsEdep1}
\end{figure}

\begin{figure}[htb]
\centering
\includegraphics[width=0.60\textwidth]{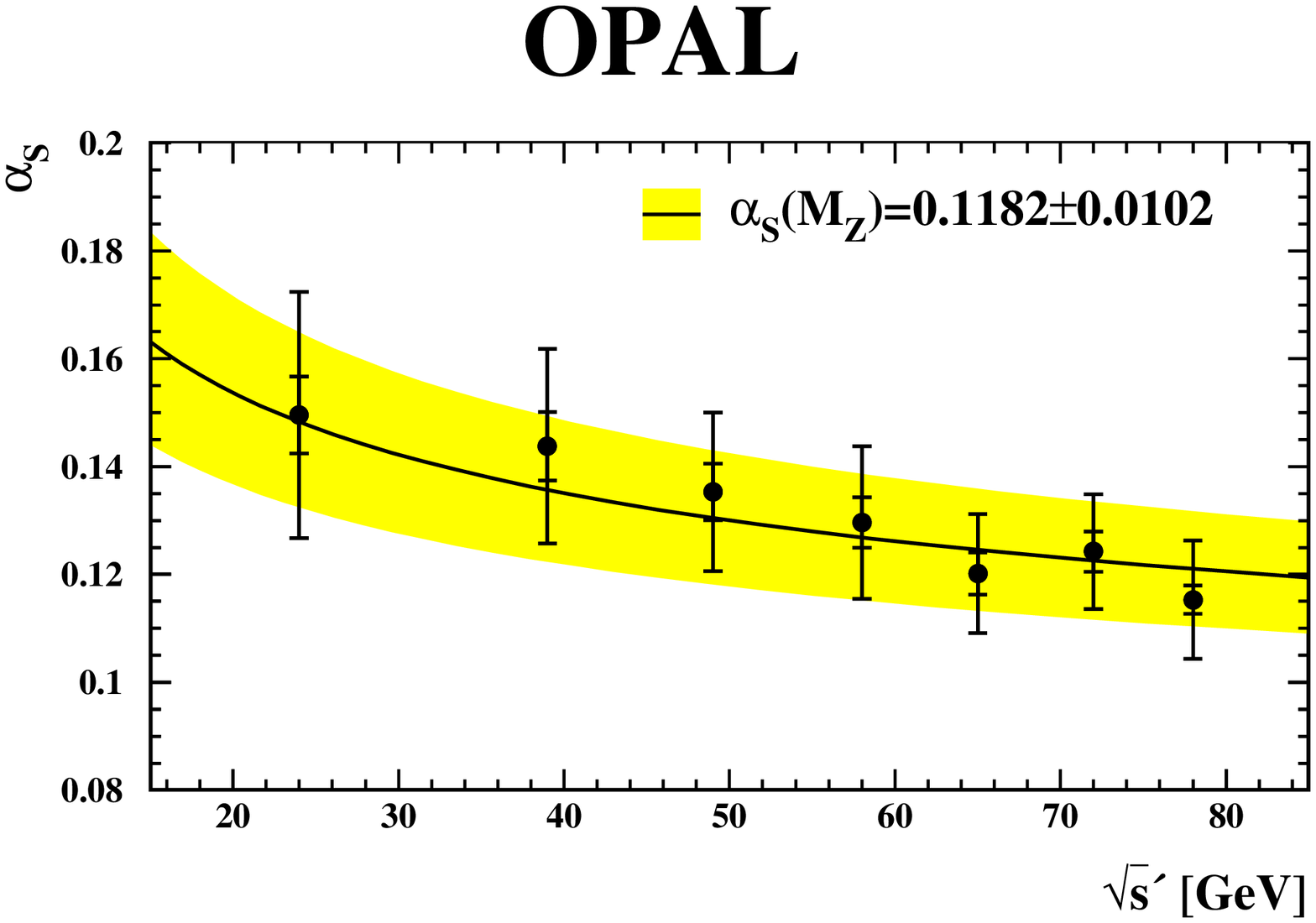}
\caption{Combined values of $\as$ from all event shape observables as
  shown in Table~\ref{VarCombinedResult}.  The curve and shaded band
  show the QCD prediction for the running of \as\ using the combined
  value of \asmz\ with total errors.}
\label{AsEdep2}
\end{figure}

The values of \as\ obtained for each observable at each energy are
used to study the energy dependence of \as\ and to obtain an overall
combined result for \asmz.  The individual values of \as\ as given in
Tables~\ref{FitResult-1}-\ref{FitResult-4} and shown in
Figure~\ref{AsEdep1} are combined taking the correlations between
their statistical and systematic errors into account using the method
described in~\cite{OPALPR404}.  The statistical covariances between
results from different observables are determined at each energy from
100 Monte Carlo subsamples with the same number of events as selected
in the data.  The experimental systematic uncertainties are assumed to
be partially correlated, i.e.\ $cov_{ij} =
\min(\sigma_i,\sigma_j)^2$.  The hadronisation and theoretical
covariances are only added to the diagonal of the total covariance
matrix.  The correlations between these uncertainties are considered
by repeating the combination procedure with different hadronisation
corrections (udsc only, \Herwig, \Ariadne) and with different
renormalisation scale parameters ($x_{\mu}=2$ and 0.5). The
systematic uncertainties for the combined value are obtained by
repeating the combination for each systematic variation. The resulting
values of \assp\ are shown in Table~\ref{VarCombinedResult} and
Figure~\ref{AsEdep2}.

\begin{figure}[htb]
\centering
\includegraphics[width=0.80\textwidth]{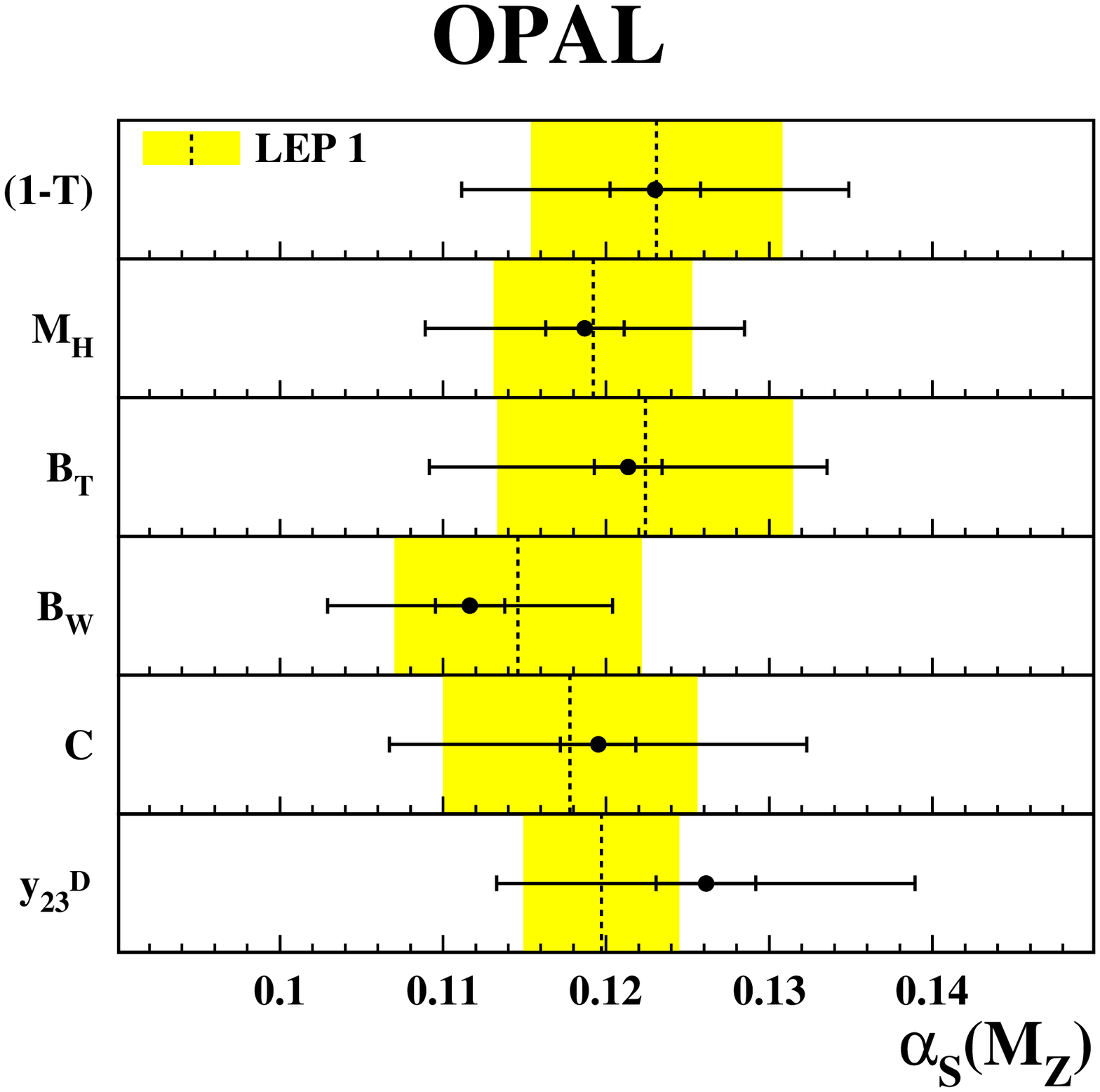}
\caption{The values of \asmz\ obtained by combining all \rsp\ samples
  as shown in Table~\ref{CombinedResult}.  The inner error bars are
  statistical, the outer error bars correspond to the total
  uncertainty. The dashed vertical lines and shaded bands show the
  LEP~1 results from OPAL~\cite{OPALPR404} using non-radiative
  events.}
\label{EcmComb1}
\end{figure}

Values of \as\ from individual observables at each energy are combined
after evolving them to $\rs=\Mz$.  In this case the results are
statistically uncorrelated.  The correlations between systematic
uncertainties are treated as explained above. The results are given in
Table~\ref{CombinedResult} and Figure~\ref{EcmComb1}.  

We also combine the combined values listed in
Table~\ref{CombinedResult} taking into account their statistical
correlations using the sum of the inverses of the individual
statistical covariance matrices at each energy point.  The result is
\begin{equation}
  \asmz=0.1182\pm 0.0015(\mathrm{stat.})\pm 0.0101(\mathrm{syst.}). 
\end{equation}
and is shown with individual errors in Table~\ref{CombinedResult}.
Figure~\ref{AsEdep2} shows the evolution of the strong coupling using
our result.  As a crosscheck on the robustness of the combination
procedure we repeat the combination using the combined results at each
energy point shown in Table~\ref{VarCombinedResult} or using all
individual results and find $\asmz=0.1183\pm0.0103$ or
$\asmz=0.1179\pm0.0103$, respectively.  

\begin{figure}[htb]
\begin{center}
\includegraphics[width=0.8\textwidth]{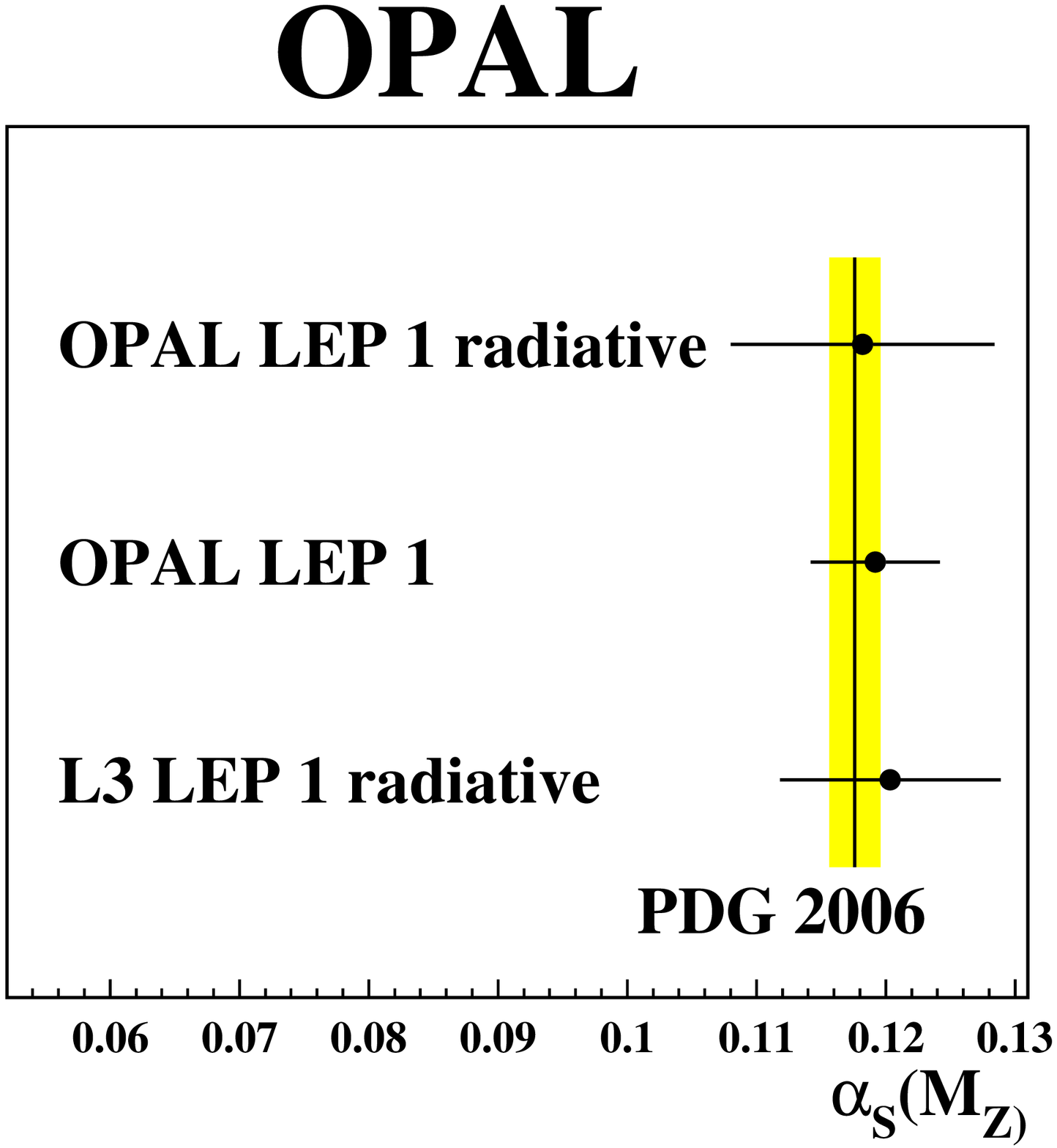}
\caption{Combined values of \asmz\ for all event shape observables and
  \rsp\ samples. The error bars show total uncertainties.  The results
  from this analyses with radiative events, from non-radiative events
  with OPAL LEP~1 data~\cite{OPALPR404} and from L3 radiative
  events~\cite{L3AsRad} are shown.  The PDG~\cite{PDG} value of \asmz\
  is also shown as the vertical line, with the total uncertainty
  corresponding to the shaded band.}
\label{EcmComb2}
\end{center}
\end{figure}

Our result is consistent within the statistical and experimental
errors with the result from OPAL using non-radiative events in LEP~1
data with the same set of observables, $\asmz=0.1192
\pm0.0002(\mathrm{stat.})\pm0.0050(\mathrm{syst.})$~\cite{OPALPR404}.  
Our result is also consistent with recent combined
values~\cite{bethke04,kluth06,kluth06b,PDG} and results from other
analyses using radiative events~\cite{L3AsRad,DELPHIas}.
Figure~\ref{EcmComb2} compares our result with the LEP~1 value for
\as\ from~\cite{OPALPR404} and an average of results from L3 using
radiative hadronic \Z\ decays~\cite{L3AsRad}\footnote{The average is
calculated using our combination procedure with the values in table~65
of~\cite{L3AsRad}.  We assume partially correlated experimental errors
and evaluate the hadronisation and theory uncertainties by repeating
the combination with simultanously changed input values.  The L3
analysis is not checked for sensitivity to the presence of massive b
quarks and thus has smaller hadronisation uncertainties. }.

The combinations of individual observables at different cms energies
yield $\chisqd\approx 1/6$.  The small values of \chisqd\ are due to
the conservative treatment of hadronisation and theoretical
uncertainties.  The values of \chisqd\ indicate consistency of the
individual results with the model of the combination including
evolution of results at different cms energies to \Mz\ before the
combination.

%%%%%%%%%%%%%%%%%%%%%%%%%%
\section{Summary}
\label{Summary}
%%%%%%%%%%%%%%%%%%%%%%%%%%

The strong coupling \as\ has been measured at reduced cms energies,
\rsp, ranging from 20~GeV to 80~GeV using event shape observables
derived from the hadronic system in radiative hadronic events.

Fits of {\cal O($\as^2$)} and NLLA QCD predictions to the six event
shape observables \thr, \mh, \bt, \bw, $C$ and $y_{23}^D$ are
performed and values of \as\ are obtained for seven values of \rsp.
Our results are consistent with the running of \as\ as predicted by
QCD.  The values at each \rsp\ are evolved to $\mu=\Mz$ and combined
for each event shape observable.  The combined value from all event
shape observables and \rsp\ values is $\asmz=0.1182\pm
0.0015(\mathrm{stat.})\pm 0.0101(\mathrm{syst.})$.

This result agrees with previous OPAL analyses with non-radiative
LEP~1 data, with a similar measurement by L3, and with recent world
average values, see figure~\ref{EcmComb2}.  Within errors, QCD is
consistent with our data sample of events with isolated FSR.  Our
result supports the assumption that the effects of high energy and
large angle FSR on event shapes in hadronic \Z\ decays can be
effectively described by QCD with a lower effective cms energy \rsp.

\section*{Acknowledgements}

We particularly wish to thank the SL Division for the efficient
operation of the LEP accelerator at all energies and for their close
cooperation with our experimental group.  In addition to the support
staff at our own institutions we are pleased to acknowledge the \\
Department of Energy, USA, \\ 
National Science Foundation, USA, \\
Particle Physics and Astronomy Research Council, UK, \\ 
Natural Sciences and Engineering Research Council, Canada, \\ 
Israel Science Foundation, administered by the Israel Academy of Science and
Humanities, \\ 
Benoziyo Center for High Energy Physics,\\ 
Japanese Ministry of Education, Culture, Sports, Science and Technology (MEXT)
and a grant under the MEXT International Science Research Program,\\
Japanese Society for the Promotion of Science (JSPS),\\ 
German Israeli Bi-national Science Foundation (GIF), \\ 
Bundesministerium f\"ur Bildung und Forschung, Germany, \\ 
National Research Council of Canada, \\ 
Hungarian Foundation for Scientific Research, OTKA T-038240, and T-042864,\\ 
The NWO/NATO Fund for Scientific Research, the Netherlands.\\

\appendix

\section{Likelihood Ratio Method}
\label{app_like}

The likelihood ratio \lrqqg\ is defined by
\begin{equation}
  \lrqqg= \frac{\lqqg}{\lqqg+\sum_i w_i L_{\mathrm{bkg},i}}
\end{equation}
where \lqqg\ and $L_{\mathrm{bkg},i}$ are the absolute likelihood
values for signal $\qqbar\gamma$ events and events from the $i$th
background process.  The background likelihood values are weighted by
$w_i$ proportional to the cross section of the $i$th background
process.

The absolute likehood values $L$ are calculated from probability
density functions (pdfs) $p_j(x_j)$ for the input variables $x_j$.
The pdfs are obtained as so-called reference histograms from simulated
signal and background samples.  For the calculation of the pdfs the
projection and correlation approximation (PCA) method~\cite{PCA} is
used.  In brief, each $x_j$ is transformed to a variable $y_j$
following a Gaussian distribution using
\begin{equation}
  y= \sqrt{2}\erf^{-1}(2F(x)-1)
\end{equation}
where $\erf^{-1}$ is the inverse error function and
$F(x)=\int_{x_{min}}^x p(x')dx'$ is the cumulative distribution of
$x$.  The likelihood $L(\vec{x})$ is then given in the PCA by
\begin{equation}
  L(\vec{x})= \frac{1}{\sqrt{|V|}} e^{-\vec{y}^T(V^{-1}-I)\vec{y}/2}
              \prod_i p_i(x_i)\;\;.
\end{equation}
V is the $n\times n$ covariance matrix of the $y_j$, $I$ is the
identity matrix and $\vec{x}$ and $\vec{y}$ are the vectors of the
$x_j$ and $y_j$.

%\newpage
%%%%%%%%%%%%%%%%%%%%%%%%%%

\clearpage
\newpage

%%% All tables

\begin{table}[htb!]
\centering
\begin{tabular}{crcccc}
\hline
{\small $E_{\gamma}$ [GeV] }& {\small Events} & {\small $\sqrt{s'}_{Mean}$[GeV] }& \multicolumn{3}{c}{{\small Background [$\%$] }} \\
\cline{4-6}
 & & & \multicolumn{2}{c}{\small Non-rad. MH } & {\small $\tau\tau$ } \\
\cline{4-6}
 & & & {\small Likelihood} & {\small Isolated tracks} & \\
\hline
10-15 & 1560 &  78.1$\pm$  1.7 &  6.0$\pm$ 0.7 &  6.2$\pm$ 0.9 &  0.9$\pm$ 0.2 \\
15-20 &  954 &  71.8$\pm$  1.9 &  3.1$\pm$ 0.5 &  4.9$\pm$ 0.8 &  1.0$\pm$ 0.3 \\
20-25 &  697 &  65.1$\pm$  2.0 &  2.6$\pm$ 0.6 &  6.3$\pm$ 1.1 &  0.9$\pm$ 0.4 \\
25-30 &  513 &  57.6$\pm$  2.3 &  5.1$\pm$ 1.1 &  7.9$\pm$ 1.4 &  1.1$\pm$ 0.5 \\
30-35 &  453 &  49.0$\pm$  2.6 &  4.5$\pm$ 1.1 &  9.6$\pm$ 1.6 &  0.7$\pm$ 0.4 \\
35-40 &  376 &  38.5$\pm$  3.5 &  5.2$\pm$ 1.2 & 13.1$\pm$ 1.9 &  0.8$\pm$ 0.5 \\
40-45 &  290 &  24.4$\pm$  5.3 & 10.4$\pm$ 2.3 & 12.9$\pm$ 1.7 &  0.8$\pm$ 0.5 \\
\hline
\end{tabular}
\caption{The number of selected events and the mean value of 
  \rsp\ for each \rsp\ subsample. The neutral hadron background
  fractions estimated by the two methods described in
  Section~\ref{BkgEstimation} are listed in the columns
  ``Non-rad. MH''.}
\label{SelectedEvents}
\end{table}

\begin{table}[htb]
\vspace*{-1cm}
\centering
\footnotesize
\begin{tabular}{|r||r|r|r|r|r|r|}
\hline
 & \multicolumn{1}{c|}{$ (1-T)$} & \multicolumn{1}{c|}{$ M_H$} & \multicolumn{1}{c|}{$ B_T$} & \multicolumn{1}{c|}{$ B_W$} & \multicolumn{1}{c|}{$ C$} & \multicolumn{1}{c|}{$ y_{23}^D$}\\
\hline\hline
 $\alpha_s( 78.1\mathrm{GeV})$  &  0.1194  &  0.1193  &  0.1144  &  0.1103  &  0.1162  &  0.1225 \\
\hline\hline
 \hline
 Statistical Error & $\pm$  0.0052 & $\pm$  0.0047 & $\pm$  0.0032 & $\pm$  0.0039 & $\pm$  0.0045 & $\pm$  0.0050\\
\hline\hline
$ \rm{Tracks+Clusters} $ & $ 0.0005$ & $-0.0005$ & $-0.0000$ & $-0.0009$ & $ 0.0002$ & $ 0.0012$\\
 \hline
$ |\cos{\theta_T}|<0.7 $ & $ 0.0096$ & $ 0.0074$ & $ 0.0059$ & $ 0.0063$ & $ 0.0067$ & $ 0.0080$\\
 \hline
$ C>5 $ & $ 0.0012$ & $ 0.0001$ & $ 0.0005$ & $-0.0004$ & $ 0.0009$ & $ 0.0006$\\
 \hline
$ \alpha^{\mathrm{iso}}_j $ & $ 0.0000$ & $ 0.0003$ & $ 0.0027$ & $ 0.0010$ & $ 0.0004$ & $-0.0012$\\
 \hline
$ \rm{Bkg\ fraction} $ & $-0.0001$ & $-0.0001$ & $-0.0001$ & $-0.0001$ & $-0.0001$ & $-0.0000$\\
 \hline
$ \rm{ECAL\ Resolution} $ & $ 0.0018$ & $ 0.0004$ & $ 0.0004$ & $ 0.0005$ & $ 0.0011$ & $-0.0005$\\
 \hline
$ \rm{Fitting\ Range} $ & $ 0.0022$ & $ 0.0005$ & $ 0.0007$ & $ 0.0016$ & $ 0.0005$ & $ 0.0005$\\
 \hline
 \hline
 Experimental Syst. & $\pm$  0.0101 & $\pm$  0.0075 & $\pm$  0.0066 & $\pm$  0.0066 & $\pm$  0.0069 & $\pm$  0.0082\\
\hline\hline
$ b - 1 s.d. $ & $-0.0005$ & $-0.0006$ & $-0.0004$ & $-0.0002$ & $-0.0006$ & $-0.0004$\\
 \hline
$ b + 1 s.d. $ & $ 0.0004$ & $ 0.0005$ & $ 0.0005$ & $ 0.0002$ & $ 0.0007$ & $ 0.0003$\\
 \hline
$ Q_0 - 1 s.d. $ & $ 0.0002$ & $-0.0004$ & $ 0.0006$ & $-0.0003$ & $ 0.0005$ & $-0.0013$\\
 \hline
$ Q_0 + 1 s.d. $ & $-0.0002$ & $ 0.0005$ & $-0.0005$ & $ 0.0003$ & $-0.0002$ & $ 0.0010$\\
 \hline
$ \sigma_q - 1 s.d. $ & $ 0.0004$ & $ 0.0003$ & $ 0.0005$ & $ 0.0003$ & $ 0.0007$ & $ 0.0007$\\
 \hline
$ \sigma_q + 1 s.d. $ & $-0.0005$ & $-0.0000$ & $-0.0005$ & $-0.0003$ & $-0.0007$ & $-0.0005$\\
 \hline
$ \rm{udsc\ only} $ & $ 0.0021$ & $-0.0001$ & $ 0.0056$ & $ 0.0023$ & $ 0.0036$ & $ 0.0065$\\
 \hline
$ \rm{Herwig\ 5.9} $ & $-0.0053$ & $-0.0046$ & $-0.0064$ & $-0.0042$ & $-0.0082$ & $-0.0078$\\
 \hline
$ \rm{Ariadne\ 4.08} $ & $ 0.0000$ & $-0.0015$ & $-0.0017$ & $-0.0001$ & $-0.0023$ & $-0.0033$\\
 \hline
 \hline
 Total Hadronisation & $\pm$  0.0057 & $\pm$  0.0049 & $\pm$  0.0087 & $\pm$  0.0048 & $\pm$  0.0093 & $\pm$  0.0108\\
\hline\hline
$ x_\mu = 0.5 $ & $-0.0051$ & $-0.0039$ & $-0.0052$ & $-0.0030$ & $-0.0053$ & $-0.0009$\\
 \hline
$ x_\mu = 2.0 $ & $ 0.0065$ & $ 0.0054$ & $ 0.0065$ & $ 0.0043$ & $ 0.0067$ & $ 0.0039$\\
 \hline
 \hline
 Total error & $+$  0.0143 & $+$  0.0115 & $+$  0.0131 & $+$  0.0100 & $+$  0.0141 & $+$  0.0150\\
 & $-$  0.0137 & $-$  0.0108 & $-$  0.0125 & $-$  0.0095 & $-$  0.0136 & $-$  0.0145\\
\hline\hline
\end{tabular}
\vspace{5mm}
\footnotesize
\begin{tabular}{|r||r|r|r|r|r|r|}
\hline
 & \multicolumn{1}{c|}{$ (1-T)$} & \multicolumn{1}{c|}{$ M_H$} & \multicolumn{1}{c|}{$ B_T$} & \multicolumn{1}{c|}{$ B_W$} & \multicolumn{1}{c|}{$ C$} & \multicolumn{1}{c|}{$ y_{23}^D$}\\
\hline\hline
 $\alpha_s( 71.8\mathrm{GeV})$  &  0.1336  &  0.1225  &  0.1304  &  0.1161  &  0.1305  &  0.1313 \\
\hline\hline
 \hline
 Statistical Error & $\pm$  0.0062 & $\pm$  0.0048 & $\pm$  0.0039 & $\pm$  0.0054 & $\pm$  0.0058 & $\pm$  0.0065\\
\hline\hline
$ \rm{Tracks+Clusters} $ & $ 0.0002$ & $ 0.0002$ & $-0.0000$ & $ 0.0001$ & $-0.0005$ & $ 0.0009$\\
 \hline
$ |\cos{\theta_T}|<0.7 $ & $ 0.0028$ & $ 0.0054$ & $ 0.0005$ & $ 0.0008$ & $-0.0024$ & $-0.0005$\\
 \hline
$ C>5 $ & $ 0.0003$ & $ 0.0010$ & $-0.0003$ & $-0.0007$ & $-0.0008$ & $-0.0011$\\
 \hline
$ \alpha^{\mathrm{iso}}_j $ & $-0.0031$ & $-0.0021$ & $-0.0022$ & $-0.0008$ & $-0.0025$ & $-0.0043$\\
 \hline
$ \rm{Bkg\ fraction} $ & $ 0.0000$ & $ 0.0001$ & $ 0.0000$ & $ 0.0001$ & $ 0.0001$ & $ 0.0000$\\
 \hline
$ \rm{ECAL\ Resolution} $ & $ 0.0015$ & $ 0.0022$ & $ 0.0014$ & $ 0.0023$ & $ 0.0007$ & $ 0.0027$\\
 \hline
$ \rm{Fitting\ Range} $ & $ 0.0020$ & $ 0.0007$ & $ 0.0007$ & $ 0.0018$ & $ 0.0004$ & $ 0.0009$\\
 \hline
 \hline
 Experimental Syst. & $\pm$  0.0049 & $\pm$  0.0064 & $\pm$  0.0028 & $\pm$  0.0032 & $\pm$  0.0037 & $\pm$  0.0054\\
\hline\hline
$ b - 1 s.d. $ & $-0.0006$ & $-0.0005$ & $-0.0005$ & $-0.0001$ & $-0.0006$ & $-0.0004$\\
 \hline
$ b + 1 s.d. $ & $ 0.0005$ & $ 0.0005$ & $ 0.0004$ & $ 0.0002$ & $ 0.0005$ & $ 0.0002$\\
 \hline
$ Q_0 - 1 s.d. $ & $ 0.0002$ & $-0.0005$ & $ 0.0007$ & $-0.0003$ & $ 0.0003$ & $-0.0017$\\
 \hline
$ Q_0 + 1 s.d. $ & $-0.0004$ & $ 0.0003$ & $-0.0007$ & $ 0.0003$ & $-0.0003$ & $ 0.0011$\\
 \hline
$ \sigma_q - 1 s.d. $ & $ 0.0004$ & $ 0.0002$ & $ 0.0005$ & $ 0.0003$ & $ 0.0005$ & $ 0.0004$\\
 \hline
$ \sigma_q + 1 s.d. $ & $-0.0005$ & $-0.0002$ & $-0.0005$ & $-0.0003$ & $-0.0005$ & $-0.0006$\\
 \hline
$ \rm{udsc\ only} $ & $ 0.0023$ & $-0.0000$ & $ 0.0061$ & $ 0.0021$ & $ 0.0033$ & $ 0.0060$\\
 \hline
$ \rm{Herwig\ 5.9} $ & $-0.0063$ & $-0.0049$ & $-0.0072$ & $-0.0041$ & $-0.0084$ & $-0.0088$\\
 \hline
$ \rm{Ariadne\ 4.08} $ & $-0.0002$ & $-0.0018$ & $-0.0017$ & $-0.0002$ & $-0.0015$ & $-0.0034$\\
 \hline
 \hline
 Total Hadronisation & $\pm$  0.0067 & $\pm$  0.0053 & $\pm$  0.0096 & $\pm$  0.0046 & $\pm$  0.0092 & $\pm$  0.0113\\
\hline\hline
$ x_\mu = 0.5 $ & $-0.0071$ & $-0.0043$ & $-0.0075$ & $-0.0034$ & $-0.0074$ & $-0.0017$\\
 \hline
$ x_\mu = 2.0 $ & $ 0.0091$ & $ 0.0060$ & $ 0.0094$ & $ 0.0049$ & $ 0.0093$ & $ 0.0049$\\
 \hline
 \hline
 Total error & $+$  0.0138 & $+$  0.0113 & $+$  0.0143 & $+$  0.0092 & $+$  0.0147 & $+$  0.0150\\
 & $-$  0.0126 & $-$  0.0105 & $-$  0.0131 & $-$  0.0085 & $-$  0.0136 & $-$  0.0142\\
\hline\hline
\end{tabular}
\caption{Values of \as\ and their errors for subsamples
  $E_{\gamma}$=10-15~GeV (upper) and 15-20~GeV (lower).}
\label{FitResult-1}
\end{table}

\begin{table}[htb]
\vspace*{-1cm}
\centering
\footnotesize
\begin{tabular}{|r||r|r|r|r|r|r|}
\hline
 & \multicolumn{1}{c|}{$ (1-T)$} & \multicolumn{1}{c|}{$ M_H$} & \multicolumn{1}{c|}{$ B_T$} & \multicolumn{1}{c|}{$ B_W$} & \multicolumn{1}{c|}{$ C$} & \multicolumn{1}{c|}{$ y_{23}^D$}\\
\hline\hline
 $\alpha_s( 65.1\mathrm{GeV})$  &  0.1236  &  0.1208  &  0.1217  &  0.1135  &  0.1242  &  0.1311 \\
\hline\hline
 \hline
 Statistical Error & $\pm$  0.0068 & $\pm$  0.0063 & $\pm$  0.0058 & $\pm$  0.0053 & $\pm$  0.0059 & $\pm$  0.0133\\
\hline\hline
$ \rm{Tracks+Clusters} $ & $-0.0011$ & $ 0.0019$ & $ 0.0020$ & $-0.0007$ & $-0.0016$ & $-0.0014$\\
 \hline
$ |\cos{\theta_T}|<0.7 $ & $ 0.0043$ & $ 0.0052$ & $ 0.0052$ & $ 0.0018$ & $ 0.0009$ & $-0.0041$\\
 \hline
$ C>5 $ & $ 0.0021$ & $ 0.0001$ & $ 0.0002$ & $ 0.0016$ & $-0.0010$ & $ 0.0009$\\
 \hline
$ \alpha^{\mathrm{iso}}_j $ & $ 0.0022$ & $ 0.0012$ & $ 0.0016$ & $ 0.0022$ & $ 0.0008$ & $ 0.0005$\\
 \hline
$ \rm{Bkg\ fraction} $ & $ 0.0001$ & $ 0.0000$ & $ 0.0000$ & $ 0.0000$ & $ 0.0000$ & $ 0.0000$\\
 \hline
$ \rm{ECAL\ Resolution} $ & $-0.0002$ & $ 0.0000$ & $ 0.0008$ & $ 0.0007$ & $ 0.0010$ & $ 0.0013$\\
 \hline
$ \rm{Fitting\ Range} $ & $ 0.0025$ & $ 0.0010$ & $ 0.0007$ & $ 0.0014$ & $ 0.0006$ & $ 0.0017$\\
 \hline
 \hline
 Experimental Syst. & $\pm$  0.0059 & $\pm$  0.0057 & $\pm$  0.0059 & $\pm$  0.0037 & $\pm$  0.0025 & $\pm$  0.0049\\
\hline\hline
$ b - 1 s.d. $ & $-0.0007$ & $-0.0006$ & $-0.0005$ & $-0.0003$ & $-0.0008$ & $-0.0002$\\
 \hline
$ b + 1 s.d. $ & $ 0.0005$ & $ 0.0007$ & $ 0.0002$ & $ 0.0003$ & $ 0.0008$ & $ 0.0004$\\
 \hline
$ Q_0 - 1 s.d. $ & $ 0.0002$ & $-0.0005$ & $ 0.0005$ & $-0.0004$ & $ 0.0004$ & $-0.0017$\\
 \hline
$ Q_0 + 1 s.d. $ & $-0.0003$ & $ 0.0004$ & $-0.0006$ & $ 0.0003$ & $-0.0002$ & $ 0.0015$\\
 \hline
$ \sigma_q - 1 s.d. $ & $ 0.0005$ & $ 0.0003$ & $ 0.0004$ & $ 0.0003$ & $ 0.0008$ & $ 0.0007$\\
 \hline
$ \sigma_q + 1 s.d. $ & $-0.0007$ & $-0.0003$ & $-0.0005$ & $-0.0004$ & $-0.0009$ & $-0.0005$\\
 \hline
$ \rm{udsc\ only} $ & $ 0.0021$ & $ 0.0001$ & $ 0.0039$ & $ 0.0025$ & $ 0.0034$ & $ 0.0062$\\
 \hline
$ \rm{Herwig\ 5.9} $ & $-0.0067$ & $-0.0051$ & $-0.0060$ & $-0.0057$ & $-0.0096$ & $-0.0099$\\
 \hline
$ \rm{Ariadne\ 4.08} $ & $-0.0007$ & $-0.0025$ & $-0.0007$ & $-0.0009$ & $-0.0027$ & $-0.0040$\\
 \hline
 \hline
 Total Hadronisation & $\pm$  0.0071 & $\pm$  0.0057 & $\pm$  0.0072 & $\pm$  0.0063 & $\pm$  0.0106 & $\pm$  0.0125\\
\hline\hline
$ x_\mu = 0.5 $ & $-0.0057$ & $-0.0042$ & $-0.0061$ & $-0.0034$ & $-0.0064$ & $-0.0014$\\
 \hline
$ x_\mu = 2.0 $ & $ 0.0073$ & $ 0.0058$ & $ 0.0076$ & $ 0.0048$ & $ 0.0081$ & $ 0.0048$\\
 \hline
 \hline
 Total error & $+$  0.0136 & $+$  0.0117 & $+$  0.0134 & $+$  0.0102 & $+$  0.0148 & $+$  0.0195\\
 & $-$  0.0128 & $-$  0.0111 & $-$  0.0126 & $-$  0.0096 & $-$  0.0140 & $-$  0.0190\\
\hline\hline
\end{tabular}
\vspace{5mm}
\footnotesize
\begin{tabular}{|r||r|r|r|r|r|r|}
\hline
 & \multicolumn{1}{c|}{$ (1-T)$} & \multicolumn{1}{c|}{$ M_H$} & \multicolumn{1}{c|}{$ B_T$} & \multicolumn{1}{c|}{$ B_W$} & \multicolumn{1}{c|}{$ C$} & \multicolumn{1}{c|}{$ y_{23}^D$}\\
\hline\hline
 $\alpha_s( 57.6\mathrm{GeV})$  &  0.1378  &  0.1396  &  0.1327  &  0.1194  &  0.1284  &  0.1407 \\
\hline\hline
 \hline
 Statistical Error & $\pm$  0.0085 & $\pm$  0.0094 & $\pm$  0.0072 & $\pm$  0.0064 & $\pm$  0.0063 & $\pm$  0.0091\\
\hline\hline
$ \rm{Tracks+Clusters} $ & $ 0.0004$ & $ 0.0022$ & $-0.0008$ & $ 0.0005$ & $ 0.0039$ & $-0.0013$\\
 \hline
$ |\cos{\theta_T}|<0.7 $ & $ 0.0065$ & $ 0.0101$ & $ 0.0078$ & $ 0.0054$ & $ 0.0083$ & $ 0.0056$\\
 \hline
$ C>5 $ & $-0.0003$ & $ 0.0020$ & $ 0.0013$ & $ 0.0013$ & $ 0.0005$ & $ 0.0009$\\
 \hline
$ \alpha^{\mathrm{iso}}_j $ & $-0.0010$ & $-0.0052$ & $ 0.0004$ & $-0.0004$ & $ 0.0001$ & $-0.0007$\\
 \hline
$ \rm{Bkg\ fraction} $ & $ 0.0000$ & $ 0.0000$ & $ 0.0000$ & $ 0.0000$ & $ 0.0000$ & $ 0.0000$\\
 \hline
$ \rm{ECAL\ Resolution} $ & $ 0.0032$ & $ 0.0051$ & $ 0.0035$ & $ 0.0021$ & $ 0.0010$ & $-0.0013$\\
 \hline
$ \rm{Fitting\ Range} $ & $ 0.0036$ & $ 0.0006$ & $ 0.0014$ & $ 0.0020$ & $ 0.0011$ & $ 0.0010$\\
 \hline
 \hline
 Experimental Syst. & $\pm$  0.0082 & $\pm$  0.0128 & $\pm$  0.0088 & $\pm$  0.0063 & $\pm$  0.0093 & $\pm$  0.0061\\
\hline\hline
$ b - 1 s.d. $ & $-0.0009$ & $-0.0004$ & $-0.0005$ & $-0.0003$ & $-0.0010$ & $-0.0006$\\
 \hline
$ b + 1 s.d. $ & $ 0.0006$ & $ 0.0004$ & $ 0.0004$ & $ 0.0004$ & $ 0.0009$ & $ 0.0005$\\
 \hline
$ Q_0 - 1 s.d. $ & $ 0.0002$ & $-0.0008$ & $ 0.0006$ & $-0.0005$ & $ 0.0005$ & $-0.0023$\\
 \hline
$ Q_0 + 1 s.d. $ & $-0.0005$ & $ 0.0005$ & $-0.0009$ & $ 0.0003$ & $-0.0004$ & $ 0.0016$\\
 \hline
$ \sigma_q - 1 s.d. $ & $ 0.0005$ & $ 0.0002$ & $ 0.0005$ & $ 0.0006$ & $ 0.0009$ & $ 0.0006$\\
 \hline
$ \sigma_q + 1 s.d. $ & $-0.0009$ & $-0.0002$ & $-0.0005$ & $-0.0004$ & $-0.0010$ & $-0.0007$\\
 \hline
$ \rm{udsc\ only} $ & $ 0.0024$ & $-0.0001$ & $ 0.0042$ & $ 0.0033$ & $ 0.0040$ & $ 0.0063$\\
 \hline
$ \rm{Herwig\ 5.9} $ & $-0.0076$ & $-0.0039$ & $-0.0072$ & $-0.0066$ & $-0.0101$ & $-0.0113$\\
 \hline
$ \rm{Ariadne\ 4.08} $ & $-0.0011$ & $-0.0011$ & $-0.0012$ & $-0.0013$ & $-0.0032$ & $-0.0049$\\
 \hline
 \hline
 Total Hadronisation & $\pm$  0.0081 & $\pm$  0.0041 & $\pm$  0.0085 & $\pm$  0.0075 & $\pm$  0.0114 & $\pm$  0.0140\\
\hline\hline
$ x_\mu = 0.5 $ & $-0.0079$ & $-0.0063$ & $-0.0078$ & $-0.0042$ & $-0.0072$ & $-0.0023$\\
 \hline
$ x_\mu = 2.0 $ & $ 0.0101$ & $ 0.0087$ & $ 0.0098$ & $ 0.0058$ & $ 0.0090$ & $ 0.0063$\\
 \hline
 \hline
 Total error & $+$  0.0175 & $+$  0.0186 & $+$  0.0172 & $+$  0.0130 & $+$  0.0183 & $+$  0.0189\\
 & $-$  0.0164 & $-$  0.0176 & $-$  0.0162 & $-$  0.0124 & $-$  0.0175 & $-$  0.0180\\
\hline\hline
\end{tabular}
\caption{Values of \as\ and their errors for subsamples 
  $E_{\gamma}=20-25$~GeV (upper) and $25-30$~GeV (lower).}
\label{FitResult-2}
\end{table}

\begin{table}[htb]
\vspace*{-1cm}
\centering
\footnotesize
\begin{tabular}{|r||r|r|r|r|r|r|}
\hline
 & \multicolumn{1}{c|}{$ (1-T)$} & \multicolumn{1}{c|}{$ M_H$} & \multicolumn{1}{c|}{$ B_T$} & \multicolumn{1}{c|}{$ B_W$} & \multicolumn{1}{c|}{$ C$} & \multicolumn{1}{c|}{$ y_{23}^D$}\\
\hline\hline
 $\alpha_s( 49.0\mathrm{GeV})$  &  0.1373  &  0.1359  &  0.1413  &  0.1269  &  0.1356  &  0.1440 \\
\hline\hline
 \hline
 Statistical Error & $\pm$  0.0105 & $\pm$  0.0098 & $\pm$  0.0087 & $\pm$  0.0069 & $\pm$  0.0089 & $\pm$  0.0117\\
\hline\hline
$ \rm{Tracks+Clusters} $ & $ 0.0022$ & $ 0.0007$ & $ 0.0032$ & $-0.0003$ & $ 0.0008$ & $-0.0012$\\
 \hline
$ |\cos{\theta_T}|<0.7 $ & $ 0.0029$ & $ 0.0039$ & $ 0.0004$ & $ 0.0012$ & $-0.0001$ & $-0.0000$\\
 \hline
$ C>5 $ & $-0.0010$ & $-0.0038$ & $-0.0017$ & $-0.0001$ & $-0.0049$ & $-0.0018$\\
 \hline
$ \alpha^{\mathrm{iso}}_j $ & $ 0.0024$ & $ 0.0024$ & $ 0.0007$ & $ 0.0017$ & $ 0.0013$ & $ 0.0046$\\
 \hline
$ \rm{Bkg\ fraction} $ & $ 0.0001$ & $ 0.0000$ & $ 0.0001$ & $ 0.0000$ & $ 0.0000$ & $ 0.0001$\\
 \hline
$ \rm{ECAL\ Resolution} $ & $-0.0003$ & $ 0.0010$ & $-0.0003$ & $ 0.0009$ & $-0.0000$ & $-0.0005$\\
 \hline
$ \rm{Fitting\ Range} $ & $ 0.0027$ & $ 0.0013$ & $ 0.0009$ & $ 0.0016$ & $ 0.0009$ & $ 0.0020$\\
 \hline
 \hline
 Experimental Syst. & $\pm$  0.0053 & $\pm$  0.0062 & $\pm$  0.0038 & $\pm$  0.0028 & $\pm$  0.0052 & $\pm$  0.0055\\
\hline\hline
$ b - 1 s.d. $ & $-0.0005$ & $-0.0009$ & $-0.0006$ & $-0.0005$ & $-0.0009$ & $-0.0008$\\
 \hline
$ b + 1 s.d. $ & $ 0.0005$ & $ 0.0008$ & $ 0.0003$ & $ 0.0005$ & $ 0.0007$ & $ 0.0002$\\
 \hline
$ Q_0 - 1 s.d. $ & $ 0.0003$ & $-0.0006$ & $ 0.0006$ & $-0.0003$ & $ 0.0005$ & $-0.0019$\\
 \hline
$ Q_0 + 1 s.d. $ & $-0.0005$ & $ 0.0005$ & $-0.0012$ & $ 0.0004$ & $-0.0005$ & $ 0.0017$\\
 \hline
$ \sigma_q - 1 s.d. $ & $ 0.0005$ & $ 0.0005$ & $ 0.0004$ & $ 0.0007$ & $ 0.0007$ & $ 0.0007$\\
 \hline
$ \sigma_q + 1 s.d. $ & $-0.0006$ & $-0.0006$ & $-0.0008$ & $-0.0007$ & $-0.0009$ & $-0.0006$\\
 \hline
$ \rm{udsc\ only} $ & $ 0.0023$ & $ 0.0002$ & $ 0.0039$ & $ 0.0050$ & $ 0.0038$ & $ 0.0060$\\
 \hline
$ \rm{Herwig\ 5.9} $ & $-0.0083$ & $-0.0090$ & $-0.0080$ & $-0.0083$ & $-0.0123$ & $-0.0114$\\
 \hline
$ \rm{Ariadne\ 4.08} $ & $-0.0009$ & $-0.0041$ & $-0.0011$ & $-0.0024$ & $-0.0039$ & $-0.0056$\\
 \hline
 \hline
 Total Hadronisation & $\pm$  0.0087 & $\pm$  0.0099 & $\pm$  0.0091 & $\pm$  0.0101 & $\pm$  0.0135 & $\pm$  0.0142\\
\hline\hline
$ x_\mu = 0.5 $ & $-0.0076$ & $-0.0058$ & $-0.0092$ & $-0.0054$ & $-0.0081$ & $-0.0008$\\
 \hline
$ x_\mu = 2.0 $ & $ 0.0097$ & $ 0.0081$ & $ 0.0117$ & $ 0.0072$ & $ 0.0102$ & $ 0.0056$\\
 \hline
 \hline
 Total error & $+$  0.0176 & $+$  0.0173 & $+$  0.0176 & $+$  0.0144 & $+$  0.0198 & $+$  0.0201\\
 & $-$  0.0165 & $-$  0.0163 & $-$  0.0160 & $-$  0.0136 & $-$  0.0188 & $-$  0.0193\\
\hline\hline
\end{tabular}
\vspace{5mm}
\footnotesize
\begin{tabular}{|r||r|r|r|r|r|r|}
\hline
 & \multicolumn{1}{c|}{$ (1-T)$} & \multicolumn{1}{c|}{$ M_H$} & \multicolumn{1}{c|}{$ B_T$} & \multicolumn{1}{c|}{$ B_W$} & \multicolumn{1}{c|}{$ C$} & \multicolumn{1}{c|}{$ y_{23}^D$}\\
\hline\hline
 $\alpha_s( 38.5\mathrm{GeV})$  &  0.1474  &  0.1374  &  0.1451  &  0.1415  &  0.1421  &  0.1496 \\
\hline\hline
 \hline
 Statistical Error & $\pm$  0.0125 & $\pm$  0.0112 & $\pm$  0.0088 & $\pm$  0.0113 & $\pm$  0.0113 & $\pm$  0.0101\\
\hline\hline
$ \rm{Tracks+Clusters} $ & $ 0.0024$ & $ 0.0019$ & $ 0.0006$ & $ 0.0001$ & $ 0.0049$ & $-0.0010$\\
 \hline
$ |\cos{\theta_T}|<0.7 $ & $ 0.0026$ & $ 0.0059$ & $ 0.0034$ & $ 0.0061$ & $ 0.0050$ & $ 0.0022$\\
 \hline
$ C>5 $ & $ 0.0042$ & $ 0.0038$ & $ 0.0018$ & $ 0.0037$ & $ 0.0052$ & $ 0.0040$\\
 \hline
$ \alpha^{\mathrm{iso}}_j $ & $ 0.0005$ & $-0.0007$ & $-0.0004$ & $ 0.0043$ & $ 0.0014$ & $ 0.0026$\\
 \hline
$ \rm{Bkg\ fraction} $ & $ 0.0003$ & $ 0.0003$ & $ 0.0002$ & $ 0.0002$ & $ 0.0003$ & $ 0.0004$\\
 \hline
$ \rm{ECAL\ Resolution} $ & $ 0.0019$ & $ 0.0025$ & $ 0.0003$ & $ 0.0035$ & $ 0.0039$ & $ 0.0055$\\
 \hline
$ \rm{Fitting\ Range} $ & $ 0.0033$ & $ 0.0009$ & $ 0.0008$ & $ 0.0013$ & $ 0.0023$ & $ 0.0008$\\
 \hline
 \hline
 Experimental Syst. & $\pm$  0.0067 & $\pm$  0.0077 & $\pm$  0.0040 & $\pm$  0.0092 & $\pm$  0.0099 & $\pm$  0.0077\\
\hline\hline
$ b - 1 s.d. $ & $-0.0009$ & $-0.0007$ & $-0.0007$ & $-0.0004$ & $-0.0007$ & $-0.0005$\\
 \hline
$ b + 1 s.d. $ & $ 0.0009$ & $ 0.0006$ & $ 0.0006$ & $ 0.0005$ & $ 0.0005$ & $ 0.0004$\\
 \hline
$ Q_0 - 1 s.d. $ & $ 0.0006$ & $-0.0008$ & $ 0.0011$ & $-0.0008$ & $ 0.0007$ & $-0.0021$\\
 \hline
$ Q_0 + 1 s.d. $ & $-0.0005$ & $ 0.0008$ & $-0.0014$ & $ 0.0006$ & $-0.0007$ & $ 0.0018$\\
 \hline
$ \sigma_q - 1 s.d. $ & $ 0.0013$ & $ 0.0003$ & $ 0.0008$ & $ 0.0006$ & $ 0.0010$ & $ 0.0005$\\
 \hline
$ \sigma_q + 1 s.d. $ & $-0.0009$ & $-0.0002$ & $-0.0008$ & $-0.0004$ & $-0.0007$ & $-0.0006$\\
 \hline
$ \rm{udsc\ only} $ & $ 0.0042$ & $ 0.0001$ & $ 0.0060$ & $ 0.0036$ & $ 0.0038$ & $ 0.0064$\\
 \hline
$ \rm{Herwig\ 5.9} $ & $-0.0150$ & $-0.0096$ & $-0.0105$ & $-0.0107$ & $-0.0125$ & $-0.0127$\\
 \hline
$ \rm{Ariadne\ 4.08} $ & $-0.0042$ & $-0.0036$ & $-0.0028$ & $-0.0025$ & $-0.0030$ & $-0.0055$\\
 \hline
 \hline
 Total Hadronisation & $\pm$  0.0162 & $\pm$  0.0103 & $\pm$  0.0125 & $\pm$  0.0116 & $\pm$  0.0135 & $\pm$  0.0154\\
\hline\hline
$ x_\mu = 0.5 $ & $-0.0093$ & $-0.0055$ & $-0.0097$ & $-0.0072$ & $-0.0089$ & $-0.0012$\\
 \hline
$ x_\mu = 2.0 $ & $ 0.0120$ & $ 0.0079$ & $ 0.0124$ & $ 0.0097$ & $ 0.0114$ & $ 0.0063$\\
 \hline
 \hline
 Total error & $+$  0.0247 & $+$  0.0188 & $+$  0.0201 & $+$  0.0210 & $+$  0.0232 & $+$  0.0210\\
 & $-$  0.0235 & $-$  0.0179 & $-$  0.0186 & $-$  0.0199 & $-$  0.0221 & $-$  0.0200\\
\hline\hline
\end{tabular}
\centering
\caption{Values of \as\ and their errors for subsamples 
  $E_{\gamma}=30-35$~GeV (upper) and $35-40$~GeV (lower).}
\label{FitResult-3}
\end{table}

\begin{table}[htb]
\centering
\vspace*{-1.4cm}
\footnotesize
\begin{tabular}{|r||r|r|r|r|r|r|}
\hline
 & \multicolumn{1}{c|}{$ (1-T)$} & \multicolumn{1}{c|}{$ M_H$} & \multicolumn{1}{c|}{$ B_T$} & \multicolumn{1}{c|}{$ B_W$} & \multicolumn{1}{c|}{$ C$} & \multicolumn{1}{c|}{$ y_{23}^D$}\\
\hline\hline
 $\alpha_s( 24.4\mathrm{GeV})$  &  0.1569  &  0.1524  &  0.1552  &  0.1433  &  0.1406  &  0.1612 \\
\hline\hline
 \hline
 Statistical Error & $\pm$  0.0252 & $\pm$  0.0117 & $\pm$  0.0115 & $\pm$  0.0101 & $\pm$  0.0112 & $\pm$  0.0181\\
\hline\hline
$ \rm{Tracks+Clusters} $ & $ 0.0038$ & $ 0.0015$ & $ 0.0060$ & $-0.0021$ & $ 0.0080$ & $-0.0074$\\
 \hline
$ |\cos{\theta_T}|<0.7 $ & $ 0.0001$ & $ 0.0008$ & $-0.0027$ & $ 0.0027$ & $ 0.0013$ & $-0.0008$\\
 \hline
$ C>5 $ & $ 0.0037$ & $-0.0001$ & $-0.0036$ & $-0.0022$ & $-0.0010$ & $-0.0084$\\
 \hline
$ \alpha^{\mathrm{iso}}_j $ & $ 0.0110$ & $ 0.0056$ & $ 0.0003$ & $ 0.0023$ & $ 0.0060$ & $ 0.0005$\\
 \hline
$ \rm{Bkg\ fraction} $ & $ 0.0023$ & $ 0.0017$ & $ 0.0018$ & $ 0.0015$ & $ 0.0020$ & $ 0.0031$\\
 \hline
$ \rm{ECAL\ Resolution} $ & $-0.0035$ & $-0.0053$ & $-0.0039$ & $-0.0013$ & $-0.0025$ & $-0.0057$\\
 \hline
$ \rm{Fitting\ Range} $ & $ 0.0035$ & $ 0.0027$ & $ 0.0018$ & $ 0.0017$ & $ 0.0020$ & $ 0.0018$\\
 \hline
 \hline
 Experimental Syst. & $\pm$  0.0134 & $\pm$  0.0085 & $\pm$  0.0088 & $\pm$  0.0054 & $\pm$  0.0109 & $\pm$  0.0131\\
\hline\hline
$ b - 1 s.d. $ & $-0.0007$ & $-0.0014$ & $-0.0007$ & $-0.0013$ & $-0.0012$ & $-0.0013$\\
 \hline
$ b + 1 s.d. $ & $ 0.0015$ & $ 0.0017$ & $ 0.0009$ & $ 0.0011$ & $ 0.0012$ & $ 0.0006$\\
 \hline
$ Q_0 - 1 s.d. $ & $ 0.0010$ & $-0.0010$ & $ 0.0023$ & $-0.0009$ & $ 0.0010$ & $-0.0039$\\
 \hline
$ Q_0 + 1 s.d. $ & $-0.0008$ & $ 0.0004$ & $-0.0029$ & $ 0.0000$ & $-0.0010$ & $ 0.0015$\\
 \hline
$ \sigma_q - 1 s.d. $ & $ 0.0014$ & $ 0.0010$ & $ 0.0011$ & $ 0.0017$ & $ 0.0012$ & $ 0.0011$\\
 \hline
$ \sigma_q + 1 s.d. $ & $-0.0010$ & $-0.0009$ & $-0.0010$ & $-0.0018$ & $-0.0013$ & $-0.0016$\\
 \hline
$ \rm{udsc\ only} $ & $ 0.0075$ & $ 0.0053$ & $ 0.0140$ & $ 0.0159$ & $ 0.0150$ & $ 0.0168$\\
 \hline
$ \rm{Herwig\ 5.9} $ & $-0.0212$ & $-0.0080$ & $-0.0134$ & $-0.0126$ & $-0.0103$ & $-0.0193$\\
 \hline
$ \rm{Ariadne\ 4.08} $ & $-0.0082$ & $-0.0056$ & $-0.0040$ & $-0.0045$ & $-0.0050$ & $-0.0114$\\
 \hline
 \hline
 Total Hadronisation & $\pm$  0.0240 & $\pm$  0.0113 & $\pm$  0.0200 & $\pm$  0.0209 & $\pm$  0.0190 & $\pm$  0.0283\\
\hline\hline
$ x_\mu = 0.5 $ & $-0.0104$ & $-0.0085$ & $-0.0116$ & $-0.0082$ & $-0.0088$ & $-0.0024$\\
 \hline
$ x_\mu = 2.0 $ & $ 0.0137$ & $ 0.0115$ & $ 0.0151$ & $ 0.0108$ & $ 0.0112$ & $ 0.0084$\\
 \hline
 \hline
 Total error & $+$  0.0397 & $+$  0.0216 & $+$  0.0289 & $+$  0.0262 & $+$  0.0270 & $+$  0.0371\\
 & $-$  0.0387 & $-$  0.0202 & $-$  0.0273 & $-$  0.0252 & $-$  0.0261 & $-$  0.0362\\
\hline\hline
\end{tabular}
\caption{Values of \as\ and their errors for subsample 
  $E_{\gamma}=40-45$~GeV.}
\label{FitResult-4}
\end{table}

\begin{table}[htb]
\centering
\small
\begin{tabular}{|c|rrrrrrr|}
\hline
\hline
\small \rsp\ [GeV] & 78.1  & 71.8  & 65.1  & 57.6  & 49.0  & 38.5  & 24.4  \\
\hline
$\as(\rsp)$ & 0.1153  & 0.1242  & 0.1201  & 0.1296  & 0.1353  & 0.1438  & 0.1496  \\
\hline
Statistical & 0.0026  & 0.0037  & 0.0039  & 0.0047  & 0.0053  & 0.0064  & 0.0071  \\
Experimental & 0.0068  & 0.0036  & 0.0040  & 0.0069  & 0.0039  & 0.0063  & 0.0077  \\
Hadronisation & 0.0062  & 0.0065  & 0.0072  & 0.0085  & 0.0100  & 0.0122  & 0.0166  \\
Theory & 0.0053  & 0.0067  & 0.0063  & 0.0076  & 0.0086  & 0.0099  & 0.0117  \\
\hline
\hline
\end{tabular}
\caption{Combined values of $\as(\rsp)$ and their errors from all
 event shape variables.}
\label{VarCombinedResult}
\end{table}

\begin{table}[htb]
 \centering
\small
\begin{tabular}{|c|rrrrrr|r|}
\hline
\hline
               & $(1-T)$  & $M_H$  & $B_T$  & $B_W$  & $C$  & $y_{23}^D$  & Combined  \\
\hline
$\alpha_s(M_Z)$ & 0.1230  & 0.1187  & 0.1214  & 0.1117  & 0.1195  & 0.1261  & 0.1182 \\
\hline
Statistical & 0.0028  & 0.0024  & 0.0021  & 0.0021  & 0.0023  & 0.0031  & 0.0015  \\
Experimental & 0.0050  & 0.0054  & 0.0037  & 0.0033  & 0.0040  & 0.0049  & 0.0038  \\
Hadronisation & 0.0071  & 0.0052  & 0.0080  & 0.0061  & 0.0092  & 0.0105  & 0.0070  \\
Theory & 0.0076  & 0.0059  & 0.0081  & 0.0049  & 0.0076  & 0.0045  & 0.0062  \\
\hline
\hline
\end{tabular}
\caption{Combined values of \asmz\ and their errors from all photon
  energy subsamples for a given observable.  The final combined value
  of \asmz\ is also shown.}
\label{CombinedResult}
\end{table}

\clearpage
\newpage

%%% All figures

\end{document}